\documentclass{aa}
\usepackage[varg]{txfonts}

\usepackage[colorlinks=true, linkcolor=blue, urlcolor=blue, citecolor=blue, anchorcolor=blue]{hyperref}
            
\usepackage{natbib}
\bibpunct{(}{)}{;}{a}{}{,} 

\usepackage{graphicx}
\usepackage{lscape}

\usepackage[normalem]{ulem}
\begin{document}

\title{Polarimetric differential imaging with VLT/NACO}
\subtitle{A comprehensive PDI pipeline for NACO data (PIPPIN)}

\author{
S. de Regt\inst{\ref{inst1}} \and
C. Ginski\inst{\ref{inst1},\ref{inst2},\ref{inst3}} \and
M. A. Kenworthy\inst{\ref{inst1}} \and
C. Caceres\inst{\ref{inst4},\ref{inst5}} \and
A. Garufi\inst{\ref{inst6}} \and
T. M. Gledhill\inst{\ref{inst7}} \and
A. S. Hales\inst{\ref{inst8},\ref{inst9}} \and
N. Huelamo\inst{\ref{inst10}} \and
\'A. K\'osp\'al\inst{\ref{inst11},\ref{inst12},\ref{inst13},\ref{inst14}} \and
M. A. Millar-Blanchaer\inst{\ref{inst15}} \and
S. P\'erez\inst{\ref{inst16},\ref{inst17},\ref{inst18}} \and
M. R. Schreiber\inst{\ref{inst5},\ref{inst19}}
}
\institute{
Leiden Observatory, Leiden University, P.O. Box 9513, 2300 RA, Leiden, The Netherlands\\\email{regt@strw.leidenuniv.nl}\label{inst1} \and
School of Natural Sciences, Center for Astronomy, University of Galway, Galway, H91 CF50, Ireland\label{inst2} \and
Anton Pannekoek Institute for Astronomy, University of Amsterdam, Science Park 904, 1098XH Amsterdam, The Netherlands\label{inst3} \and
Instituto de Astrof\'isica, Facultad de Ciencias Exactas, Universidad Andres Bello, Av. Fern\'andez Concha 700, Santiago, Chile\label{inst4} \and
N\'ucleo Milenio de Formaci\'on Planetaria (NPF), Valpara'iso, Chile\label{inst5} \and
INAF, Osservatorio Astrofisico di Arcetri, Largo Enrico Fermi 5, I-50125 Firenze, Italy\label{inst6} \and
Department of Physics, Astronomy \& Mathematics, University of Hertfordshire, College Lane, Hatfield, Hertfordshire AL10 9AB, UK\label{inst7} \and
Joint ALMA Observatory, Avenida Alonso de C\'ordova 3107, Vitacura 7630355, Santiago, Chile\label{inst8} \and
National Radio Astronomy Observatory, 520 Edgemont Road, Charlottesville, VA 22903-2475, United States of America\label{inst9} \and
Centro de Astrobiolog\'{\i}a (CAB), CSIC-INTA, ESAC Campus, Camino bajo del Castillo s/n, E-28692 Villanueva de la Ca\~nada, Madrid, Spain\label{inst10} \and
Konkoly Observatory, HUN-REN Research Centre for Astronomy and Earth Sciences, Konkoly-Thege Mikl\'os \'ut 15-17, 1121 Budapest, Hungary\label{inst11} \and
CSFK, MTA Centre of Excellence, Konkoly-Thege Mikl\'os \'ut 15-17, 1121 Budapest, Hungary\label{inst12} \and
ELTE E\"otv\"os Lor\'and University, Institute of Physics and Astronomy, P\'azm\'any P\'eter s\'et\'any 1/A, 1117 Budapest, Hungary\label{inst13} \and
Max Planck Institute for Astronomy, K\"onigstuhl 17, 69117 Heidelberg, Germany\label{inst14} \and
Department of Physics, University of California, Broida Hall, Santa
Barbara, CA 93106, USA\label{inst15} \and
Departamento de Fisica, Universidad de Santiago de Chile, Av. Victor Jara 3659, Santiago, Chile\label{inst16} \and
Millennium Nucleus on Young Exoplanets and their Moons (YEMS), Santiago, Chile\label{inst17} \and
Center for Interdisciplinary Research in Astrophysics and Space Exploration (CIRAS), Universidad de Santiago, Santiago, 9170124, Chile\label{inst18} \and
Departamento de F\'isica, Universidad T\'ecnica Federico Santa Mar\'ia, Av. España 1680, Valpara\'iso, Chile\label{inst19}
}

\date{Received 26 November 2023 / Accepted 19 February 2024}

\abstract
{The observed diversity of exoplanets can possibly be traced back to the planet formation processes. Planet--disk interactions induce sub-structures in the circumstellar disk that can be revealed via scattered light observations. However, a high-contrast imaging technique such as polarimetric differential imaging (PDI) must first be applied to suppress the stellar diffraction halo.}
{In this work we present the PDI PiPelIne for NACO data (PIPPIN), which reduces the archival polarimetric observations made with the NACO instrument at the Very Large Telescope. Prior to this work, such a comprehensive pipeline to reduce polarimetric NACO data did not exist. We identify a total of 243 datasets of 57 potentially young stellar objects observed before NACO's decommissioning.}
{The PIPPIN pipeline applies various levels of instrumental polarisation correction and is capable of reducing multiple observing setups, including half-wave plate or de-rotator usage and wire-grid observations. A novel template-matching method is applied to assess the detection significance of polarised signals in the reduced data.}
{In 22 of the 57 observed targets, we detect polarised light resulting from a scattering of circumstellar dust. The detections exhibit a collection of known sub-structures, including rings, gaps, spirals, shadows, and in- or outflows of material. Since NACO was equipped with a near-infrared wavefront sensor, it made unique polarimetric observations of a number of embedded protostars. This is the first time detections of the Class I objects Elia 2-21 and YLW 16A  have been published. Alongside the outlined PIPPIN pipeline, we publish an archive of the reduced data products, thereby improving the accessibility of these data for future studies.}
{}


\maketitle

\section{Introduction} \label{Section1}

Over 5\,500 exoplanets\footnote{February 2024; \url{https://exoplanets.nasa.gov/discovery/exoplanet-catalog/}} have been discovered to date, and they are extremely diverse in terms of their masses, compositions, and distributions around their parent stars. Planet formation theories, such as the core-accretion \citep{Pollack_1996} or disk gravitational instability \citep{Boss_1997} models, must be able to explain the resulting diverse planetary systems. To investigate the formation processes, we can study the circumstellar disks that shape the planet-forming environments. Disk sub-structures, such as rings or cavities, are expected byproducts of planet formation and are indeed associated with the protoplanet-hosting PDS 70 \citep{Keppler_2018,Keppler_2019,Haffert_2019} and AB Aur systems \citep{Currie_2022}, although the evidence for AB Aur b was recently disputed by \citet{Zhou_2023}. Multi-wavelength observations trace different disk regions, including the large, millimetre-sized dust grains near the midplane \citep[e.g.][]{ALMA_partnership_2015} at longer wavelengths. Scattered light can be captured from the upper surfaces of the disk at optical and near-infrared (NIR) wavelengths and provides information about the material through the measurements of phase functions and the degree of polarised light. Since central stars are observed close to the peak of blackbody emission, a high-contrast imaging technique is employed to reveal the faint structures in their immediate vicinities. Polarimetric differential imaging \citep[PDI; ][]{Gledhill_1991,Gledhill_2001,Kuhn_2001} is especially well suited to observing the optical and NIR scattered light of a circumstellar disk. Unpolarised stellar light becomes polarised after being scattered by circumstellar dust grains, and PDI can be used to remove the stellar component, revealing the fainter polarised light structures below the diffraction halo of the star.

Several instruments, including the Subaru High-Contrast Coronagraphic Imager for Adaptive Optics \citep[HiCIAO;][]{Hodapp_2008, Suzuki_2010}, the Gemini South Gemini Planet Imager \citep[GPI;][]{Macintosh_2006, Macintosh_2014}, the Very Large Telescope (VLT) Nasmyth Adaptive optics system COude (NACO) NIR camera \citep[][]{Lenzen_2003, Rousset_2003}, and the VLT Spectro-Polarimetric High-contrast Exoplanet REsearch instrument \citep[SPHERE; ][]{Beuzit_2019}, have exploited the PDI technique to observe a large number of young stellar objects (YSOs). These instruments utilise a polarised beam-splitter to separate the incoming light into two beams with orthogonal linear polarisations. The instrumental point spread function (PSF) is unchanged for both beams, as they are recorded simultaneously. The high contrast \citep[$\sim$\,$10^{-2}$ -- $10^{-4}$;][]{Avenhaus_2018} between the faint scattered light disk and the bright stellar halo can be suppressed by subtracting measurements of the two orthogonal polarisation states. In particular, PDI is an effective imaging technique for circumstellar disks with low inclinations \citep[e.g. HD 169142, $i\approx13^\circ$ and TW Hya, $i\approx7^\circ$;][]{Hales_2006,Apai_2004,van_Boekel_2017}, for which angular differential imaging \citep[][]{Marois_2006} leads to the self-subtraction of the face-on disk's signal. 

Observations that use PDI have revealed a large number of disks with different sizes, surface brightnesses, and morphologies in scattered light. Scattered light observations trace the upper layers of a circumstellar disk since the micron-sized dust grains found there are optically thick at optical and NIR wavelengths. Hence, circumstellar disks must be flared to intercept the stellar radiation at large distances \citep{Chiang_1997, de_Boer_2016, Ginski_2016}. Transition disks with large dust-depleted inner cavities are frequently detected \citep[e.g.][]{Mayama_2012, Canovas_2013, Keppler_2018, Mauco_2020}, and the observed circumstellar disks commonly show rings at varying radii \citep[e.g.][]{Quanz_2013,Muro_Arena_2018,Avenhaus_2018}. Additionally, spiral features are frequently detected in scattered light (see Fig. 9 of \citealt{Benisty_2022}). The gas perturbations, coupled to the small grains that are traced in scattered light, are suggested to emerge from interactions with a companion or with the environment. Furthermore, the combination with sub-millimetre observations can reveal dust filtering at pressure maxima \citep[e.g.][]{Garufi_2013, Mauco_2020} and help identify fragmentation, possibly resulting from gravitational instability \citep{Weber_2023}. In scattered light imaging, the misalignment of an (un-resolved) inner disk can cast a shadow onto the outer disk \citep{Bohn_2022}. Depending on the magnitude of the misalignment, narrow shadow lanes \citep[e.g. HD 100453;][]{Benisty_2017} or wide-angle obscurations can appear \citep[e.g. HD 143006;][]{Benisty_2018}. In the case of stellar multiplicity, the geometry of the circumstellar environment can be assessed further by interpreting which stellar component is responsible for the dust illumination \citep{Weber_2023,Zurlo_2023}. Depending on the size, composition, and porosity of the small dust grains, different scattering phase functions can be measured \citep{Shen_2009, Tazaki_2016, Tazaki_2019}. By studying the dust properties in circumstellar disks, we can assess the efficiency of dust growth based on the size, composition, and porosity of the grains involved. PDI observations are not limited to Class II disks \citep{Lada_1987}: second-generation dust disks, or debris disks, are also observed with facilities such as VLT/SPHERE (e.g. HIP 79977; \citealt{Engler_2017}, HR 4796A; \citealt{Milli_2019}), Gemini South/GPI (e.g. HD 157587; \citealt{Millar_Blanchaer_2016}), and Subaru/HiCIAO (e.g. HD 32297; \citealt{Asensio_Torres_2016}).

For the most studied Class II disks \citep{Lada_1987}, the observed sub-structures are frequently explained by invoking the presence of planetary companions \citep[e.g.][]{ALMA_partnership_2015, van_der_Marel_2019, Long_2019, Asensio_Torres_2021}. The existence of sub-structures suggests that planet formation is already underway and began when the YSOs were still embedded in their natal envelopes, during the Class 0 or I phases \citep[$t<10^6\mathrm{\ yr}$;][]{Garufi_2022a}. Furthermore, measurements of the dust masses of Class II disks appear incompatible with predicted planet formation efficiencies and the masses of exoplanetary systems \citep{Manara_2018, Mulders_2021}. The higher dust masses of Class 0 and I disks determined by \citet{Tychoniec_2020} could indicate that giant planet formation commences before the protostellar envelope has dissipated \citep{Cridland_2022, Miotello_2022}. Alternatively, the accretion of material from the surrounding cloud can continually replenish the mass of the protoplanetary disk. The total mass budget available for planet formation therefore exceeds the disk mass at any given time \citep{Manara_2018, Garufi_2022a}. The two explanations put forward to solve the missing mass problem demonstrate the important role of embedded Class 0 and I objects in the formation of planets. However, the earliest YSOs are particularly difficult to observe at optical wavelengths due to their embedded nature. As a consequence, the optical wavefront sensors (WFSs) of most modern extreme adaptive optics (AO) systems do not allow for an adequate AO correction of deeply embedded YSOs. The NIR AO188 system, part of SCExAO on the Subaru telescope, is an exception as it provides AO for polarimetric imaging in the northern hemisphere. However, embedded protostars in the south were only observable to some older, ground-based instruments that were equipped with an infrared WFS. For completeness, the retired NICMOS instrument on the \textit{Hubble} Space Telescope measured the polarised light of the earliest YSOs \citep[e.g.][]{Silber_2000,Kospal_2008,Perrin_2009}, though JWST is not equipped with polarimetric capabilities.

In this work, we present a re-reduction of polarimetric archival data from NACO, the Nasmyth Adaptive Optics System (NAOS), which together with the COude Near-Infrared CAmera (CONICA) forms the NACO instrument at the VLT \citep{Lenzen_2003, Rousset_2003}. Initially installed at the Nasmyth B focus of UT4 in 2001, NACO was reinstalled at the Nasmyth A focus of UT1 from 2014 until its decommissioning in 2019. NACO operated at wavelengths between $1$ and $5\ \mu\mathrm{m,}$ and NAOS was equipped with a visible ($0.45$--$1.0\ \mu\mathrm{m}$) and infrared ($0.8$--$2.5\ \mu\mathrm{m}$) WFS, enabling observations of embedded YSOs despite their faint optical magnitudes. NACO was equipped with a Wollaston prism (and also wire grids; see Sect. \ref{sect:rotator_wiregrid}) to perform polarimetric observations, and a half-wave plate (HWP) was installed in 2003. In Sect. \ref{Section2} we describe how our PDI PiPelIne for NACO data (PIPPIN)\footnote{PIPPIN is a publicly available Python package; see: \url{https://pippin-naco.readthedocs.io} for more information.} reduces the NACO polarimetric data with the PDI technique. Section \ref{Section3} outlines a broad inspection of the PIPPIN-reduced data, and we present a novel method for assessing the detection significance of a polarised signal. Section \ref{Section5} compares the reduced data with those of the SPHERE instrument. The conclusions are summarised in Sect. \ref{Section6}, and the reduced data archive has been published online\footnote{\url{https://doi.org/10.5281/zenodo.8348803}}. 

\section{Reduction of NACO data} \label{Section2}
\subsection{Selection of polarimetric observations}
Since the polarimetric mode of NACO was not solely used to observe YSOs, we made a selection of observations of interest to this study. First, the European Southern Observatory (ESO) archive was searched for every polarimetric NACO observation classified as the SCIENCE data type. Using the object identifier and the \texttt{astroquery} Python package \citep{Ginsburg_2019}, we searched the SIMBAD archive \citep{Wenger_2000} to select any object that was ever classified as one of the following categories: (candidate) Orion variable, (candidate) Herbig Ae/Be star, (candidate) T Tauri star, or a (candidate) YSO. However, a large number of observations have unclear object identifiers. In these instances, \texttt{astroquery} was utilised to locate the object closest to the target right ascension (RA) and declination (Dec.) coordinates. In total, we find 57 candidate Class 0 - III objects that potentially exhibit polarised light from circumstellar material. As these systems were observed in multiple filters, epochs, or with different instrument setups, we find a total of 243 datasets. Table \ref{tab:systems} lists the objects of interest and information on the observation setup for each dataset. 

\subsection{PDI PiPelIne for NACO data (PIPPIN)}
A general pipeline to reduce NACO data is provided by ESO\footnote{\url{https://www.eso.org/sci/software/pipelines/naco/naco-pipe-recipes.html}}. However, this pipeline cannot reduce the polarimetric observations and thus previous works utilised custom, self-written pipelines (e.g. \citealt{Apai_2004,Quanz_2011,Canovas_2013}). The different data reduction methods could lead to inconsistent scientific results. For instance, one of the rings of HD 97048 observed by \citet{Ginski_2016} was not recovered from the same data in the earlier analysis of \citet{Quanz_2012}. Such discrepancies can be avoided by using a single, comprehensive pipeline. In this section, we describe the operation of our PIPPIN pipeline, which applies the PDI technique to polarimetric NACO observations. With the exception of an instrumental Mueller matrix model, PIPPIN largely follows the polarimetric data reduction outlined in \citet{de_Boer_2019}. For a more detailed characterisation of the instrumental polarisation of NACO, we refer to \citet{de_Boer_2014} and \citet{Millar_Blanchaer_2020}.

\subsubsection{FLATs, bad-pixel masks, and DARKs}
To correct for any variations of the detector's gain, PIPPIN performs a FLAT-fielding of the SCIENCE images. In general, internal lamp FLATs were taken for each filter and detector (i.e. S13, S27, L27, S54, and L54) that were used during the night. The polarimetric mask, which prevents the ordinary and extra-ordinary beams from overlapping, is also inserted when measuring the FLAT fields. The FLATs are DARK-subtracted and subsequently normalised by being divided by the median counts. The bad-pixel masks are generated by assessing which pixels had a non-linear response in the FLAT fields. The linearity of the pixel response is determined by comparing the FLATs observed with the internal lamp switched on ($\mathrm{FLAT}_\mathrm{on}$) to FLATs made with the lamp turned off ($\mathrm{FLAT}_\mathrm{off}$). The factor by which the pixel-counts are expected to increase is computed by dividing the median of $\mathrm{FLAT}_\mathrm{on}$ by the median of $\mathrm{FLAT}_\mathrm{off}$. Pixels were flagged when their response deviated by more than $2\sigma$ from the expected increase. Similar to the FLAT fields, the bad-pixel masks are computed for each filter and detector used throughout the night.

\subsubsection{Pre-processing}
The PIPPIN pipeline can be described in two parts: the pre-processing and the application of PDI. The pre-processing commences by reading parameters from a configuration file that allows users to customise the data reduction. The configuration file must be located in the same directory as the SCIENCE observations; otherwise, the pipeline creates a default file. Table \ref{tab:PIPPIN_keywords} outlines the parameter keywords in the configuration file along with the recognised values, descriptions, and default values. After reading the configuration parameters, PIPPIN groups observations by the utilised detector, window-size, observing ID (if requested), filter, exposure time, HWP usage, and whether the Wollaston prism or wire grids were used. Each observation is DARK-subtracted and FLAT-normalised by division. The pixels flagged in the bad-pixel mask are replaced by the median counts of the surrounding square of $5\times5$ pixels. 

To retrieve the approximate positions of the ordinary and extra-ordinary beams, PIPPIN applies a minimum-filter with a specific kernel-shape to the images. The filter consists of two squares of $3\times3$ pixels that are offset by the approximate separation of the beams, which in turn depends on the pixel scale of the utilised detector. The maximum in the filtered image yields the approximate location of the ordinary and extra-ordinary beams. This method avoids any persisting bad pixels or image artefacts such as the polarimetric mask. Subsequently, the initial guesses are used to retrieve more accurate PSF locations via a user-specified fitting method. For each beam, PIPPIN can employ a single 2D Moffat function or subtract two Moffat functions from each other to reproduce the flat top of a saturated PSF. Alternatively, the pipeline can use a maximum-counts method for the asymmetric PSFs that are encountered in the case of deeply embedded stars.

\begin{figure}[h!]
    \resizebox{\hsize}{!}{\includegraphics[width=17cm]{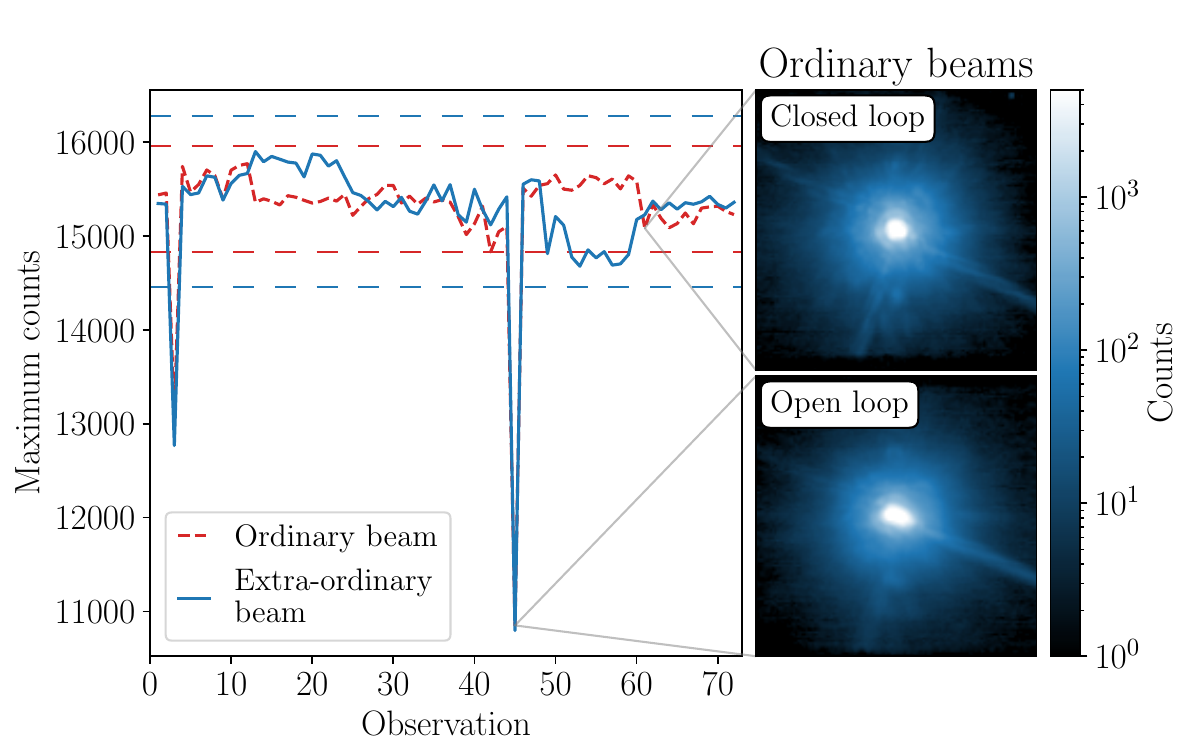}}
    \caption{Open AO-loop assessment of HD 135344B Ks-band observations. \textit{Left panel}: Maximum counts of the ordinary (red) and extra-ordinary (blue) beams. The horizontal dashed blue and red lines are the $3\sigma$ bounds for the respective beam, indicating which observations have adequate AO corrections. The \textit{upper-right panel} shows an example of an effective AO correction for the ordinary beam, and the \textit{lower-right panel} shows the blurred result of an open AO loop.}
    \label{fig:open_AO_example}
\end{figure}

The sky-subtraction is performed by subtracting two dithering positions or by subtracting the median per row of pixels. To avoid contamination from the target, a region around the fitted beam centres is excluded in the median sky-subtraction method. This region is defined with the \texttt{sky\_subtraction\_min\_offset} parameter in the configuration file. Moreover, this parameter ensures that the two dithering positions are sufficiently offset to perform a sky-subtraction. In addition, horizontal gradients are removed by a linear fit that excludes the region around the beams. The linear fit is applied to the average of five rows of pixels and a 2D Gaussian filter with $\sigma=5\ \mathrm{pixels}$ is applied to smooth out the resulting background approximation. Some observations show a distinct horizontal pattern, which can be removed by fitting each row of pixels individually and without applying a Gaussian filter. Next, the ordinary and extra-ordinary beams are cut out of the images by a user-specified crop-size. The maximum counts of the beams are evaluated with an iterative sigma-clipping to determine which observations suffered from a poor AO correction. Figure \ref{fig:open_AO_example} shows an example of the open AO-loop analysis for observations of HD 135344B in the Ks band. The left panel shows the maximum counts of the ordinary and extra-ordinary beams for each observation in red and blue, respectively. The horizontal dashed lines show the $3\sigma$-bounds used in the sigma-clip. The right panels show examples of the ordinary beams of two observations. In this work, the images presented with a blue colour map show PIPPIN-reduced data products. The upper-right panel of Fig. 
\ref{fig:open_AO_example} shows an effective AO correction and the lower-right panel shows an example of an open AO loop. In the bottom panel, we notice that the point source is blurred, likely as a result of a tilting wavefront during the integration. The resulting maximum count of the (extra-)ordinary beam is measured lower than the $3\sigma$-bound and this observation was removed. In this example, observations 3, 42, and 45 were ignored during the PDI application.

\subsubsection{Polarimetric differential imaging}\label{sect:PDI}
Polarised light can be described with the Stokes formalism and the Stokes vector:
\begin{align}
    \vec{S} &= \begin{pmatrix}
    I \\
    Q \\
    U \\
    V
    \end{pmatrix},
\end{align}
where $I$ is the total intensity, $Q$ and $U$ are intensities of the linear polarisation components and $V$ describes the circular polarisation. As NACO was not primarily designed for polarimetry, the observations suffer from instrumental polarisation ($IP$) and crosstalk effects. Reflections within the instrument can introduce polarised signal whose magnitude depends on the instrument configuration, altitude of the target object, etc. Furthermore, crosstalk between the linear and circular polarisation components reduces the polarimetric efficiency \citep{Witzel_2011}. Hence, PIPPIN employs a multi-stage correction for these effects. A first-order correction for different transmission efficiencies is to impose that the stellar flux in the ordinary ($I_\mathrm{ord}$) and extra-ordinary ($I_\mathrm{ext}$) beams are the same, as described in Appendix C of \citet{Avenhaus_2014a}. Since the PSF core is often saturated in NACO observations, PIPPIN draws multiple, user-specified annuli and computes the total fluxes within them. For each annulus $i$, the ratio between the fluxes, 
\begin{align}
    X_{\mathrm{ord/ext},i} = \frac{\sum_\mathrm{pixels}I_{\mathrm{ord},i}}{\sum_\mathrm{pixels}I_{\mathrm{ext},i}}, \label{eq:X_oe}
\end{align}
is used to scale the ordinary and extra-ordinary images as $I_\mathrm{ord}/\sqrt{X_{\mathrm{ord/ext},i}}$ and $I_\mathrm{ext}\sqrt{X_{\mathrm{ord/ext},i}}$, respectively. This method implicitly assumes that the total flux in annulus $i$ is unpolarised, thereby ignoring any polarisation induced by the interstellar medium or any intrinsic polarisation originating from an unresolved inner disk, for example. We note that this correction could overcompensate for a true disk signal if the disk is not axisymmetric and if its scattered light comprises a considerable fraction of the stellar signal.

\begin{figure}[h!]
    \resizebox{\hsize}{!}{\includegraphics[width=17cm]{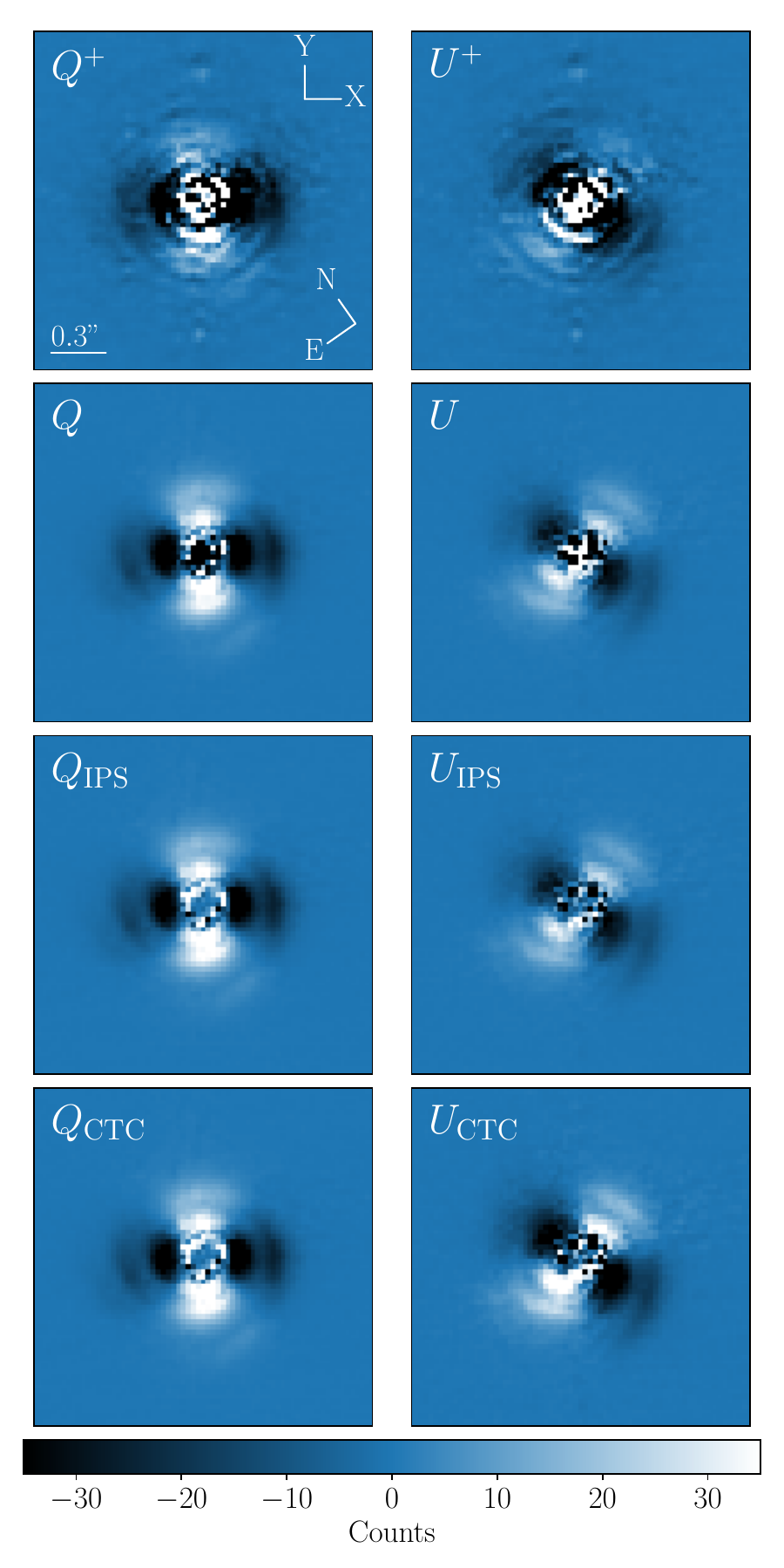}}
    \caption{Median Stokes $Q$ and $U$ images with different levels of $IP$ corrections for HD 135344B Ks-band observations. \textit{From top to bottom}: $Q^+$ and $U^+$ components after equalising the ordinary and extra-ordinary fluxes, $Q$ and $U$ resulting from the double-difference method, $Q_\mathrm{IPS}$ and $U_\mathrm{IPS}$ after subtracting the median $IP$ within an annulus, and the crosstalk-corrected $Q_\mathrm{CTC}$ and $U_\mathrm{CTC}$ components where the reduced Stokes $U$ efficiency is accounted for. The characteristic butterfly pattern is visible in each panel, and the compasses show the orientation of the detector and the sky.}
    \label{fig:QU}
\end{figure}

If the HWP has a rotation angle of $\theta=0^\circ$, the ordinary beam ($I_\mathrm{ord}$) measures light polarised in the $+Q$ direction and the extra-ordinary beam ($I_\mathrm{ext}$) measures the perpendicularly polarised light in the $-Q$ direction, both in the HWP reference frame. The $I_Q$ and $Q$ components are found by addition and subtraction of the equalised beam intensities:
\begin{align}
    I_Q &= I_\mathrm{ord} + I_\mathrm{ext}\big|_{\theta=0^\circ}, \\
    Q &= I_\mathrm{ord} - I_\mathrm{ext}\big|_{\theta=0^\circ}.
\end{align}
Measurements of the $U$ component are made by rotating the incoming beam by $45^\circ$, which means that the HWP is rotated by $\theta=22.5^\circ$. The $I_U$ and $U$ components are calculated with
\begin{align}
    I_U &= I_\mathrm{ord} + I_\mathrm{ext}\big|_{\theta=22.5^\circ}, \\
    U &= I_\mathrm{ord} - I_\mathrm{ext}\big|_{\theta=22.5^\circ}.
\end{align}
The top panels of Fig. \ref{fig:QU} show the resulting median Stokes $Q$ and $U$ images for HD 135344B. The position angle (PA) is $-35^\circ$, so that the sky is rotated anti-clockwise to the axes of the detector as is indicated by the compasses in the figure. In the $Q$ image, the positive signal aligns with the $\textit{Y}$-axis and the negative signal aligns with the $\textit{X}$-axis. The $U$ image displays a similar butterfly pattern, but rotated by $45^\circ$ since it measures different components of the disk. 

\begin{figure*}[h!]
    \centering
    \includegraphics[width=17cm]{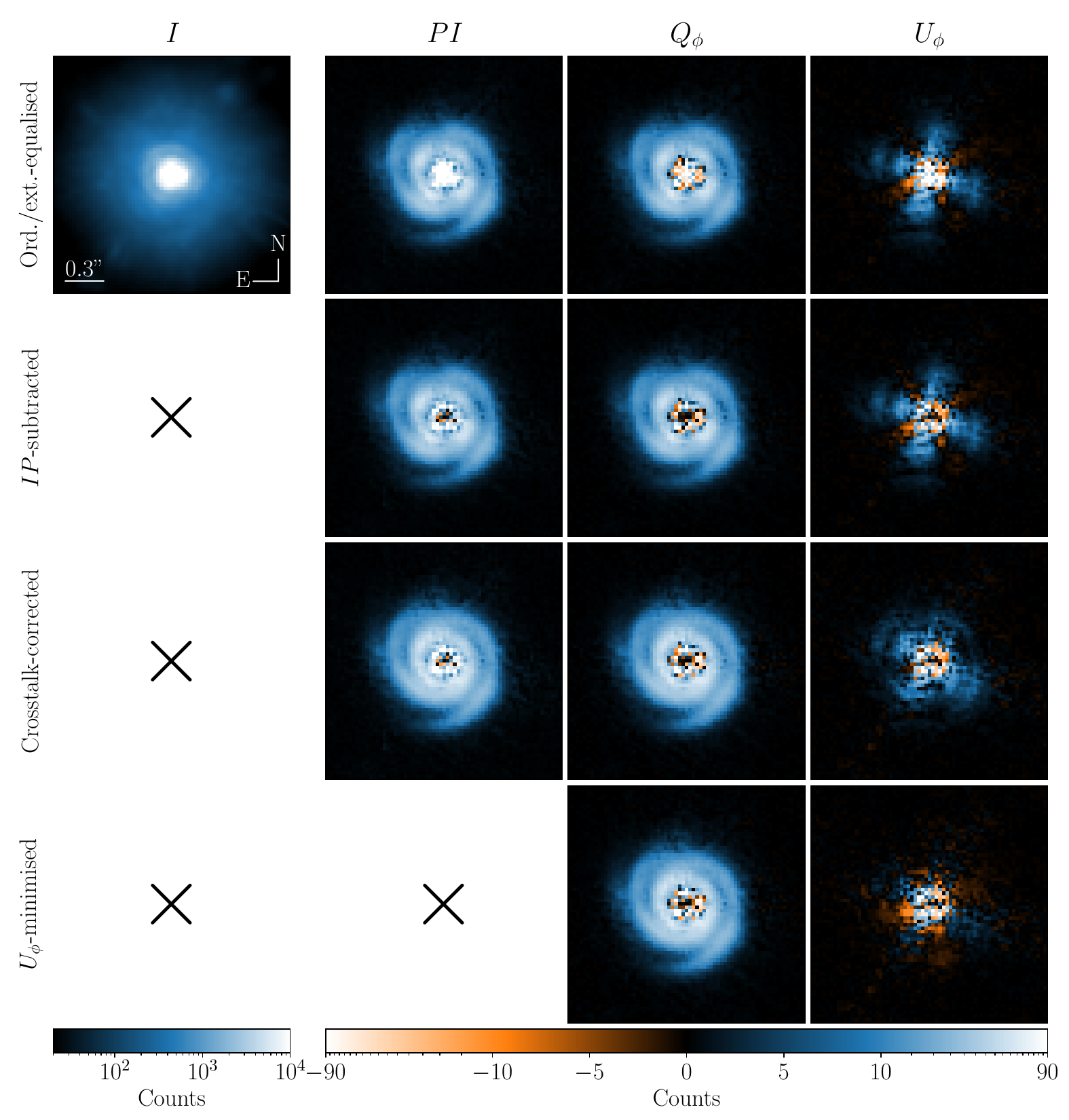}
    \caption{Final PIPPIN data products with different levels of $IP$ correction. \textit{From left to right}: Median total intensity ($I)$, polarised intensity $(PI$), and the azimuthal Stokes components $Q_\phi$ and $U_\phi$ of HD 135344B observed in the Ks band. \textit{From top to bottom}: Equalised ordinary and extra-ordinary beams, $IP$-subtracted, crosstalk-corrected, and $U_\phi$-minimised results. The total intensity is shown with a logarithmic scale from $20$ to $10^4$ counts, whereas the other panels use a linear scale from $-5$ to $+5$ counts and a logarithmic scale up to $\pm90$. The negative signal is depicted in orange, and in each image north points up and east to the left.}
    \label{fig:PI}
\end{figure*}

Instrumental polarisation introduced downstream of the HWP can be removed by recording the $-Q$ and $-U$ parameters at $\theta=45^\circ$ and $67.5^\circ$, respectively. The instrumental $Q_{IP}$ and $U_{IP}$ components are unaffected by this rotation of the HWP and contribute in the same manner as before:
\begin{alignat}{3}
    Q^+ &= Q + Q_\mathrm{IP}   &&= I_\mathrm{ord} - I_\mathrm{ext}\big|_{\theta=0^\circ}, \label{eq:Q+}\\
    Q^- &= - Q + Q_\mathrm{IP} &&= I_\mathrm{ord} - I_\mathrm{ext}\big|_{\theta=45^\circ}, \label{eq:Q-}\\
    U^+ &= U + U_\mathrm{IP}   &&= I_\mathrm{ord} - I_\mathrm{ext}\big|_{\theta=22.5^\circ}, \label{eq:U+}\\
    U^- &= - U + U_\mathrm{IP} &&= I_\mathrm{ord} - I_\mathrm{ext}\big|_{\theta=67.5^\circ}. \label{eq:U-}
\end{alignat}
Using the double-difference method \citep{Hinkley_2009,Bagnulo_2009}, we can subtract the $IP$ components:
\begin{align}
    Q &= \frac{1}{2}(Q^+ - Q^-), \label{eq:double_diff1} \\
    U &= \frac{1}{2}(U^+ - U^-). \label{eq:double_diff2}
\end{align}
Similarly, the $IP$-corrected intensities are found with the double-sum:
\begin{align}
    I_Q &= \frac{1}{2}(I_{Q^+} + I_{Q^-}), \label{eq:double_sum1} \\
    I_U &= \frac{1}{2}(I_{U^+} + I_{U^-}). \label{eq:double_sum2}
\end{align}
The total intensity is calculated as\begin{align}
    I = \frac{1}{2}(I_Q + I_U).
\end{align}
The second row of Fig. \ref{fig:QU} shows the median $Q$ and $U$ images resulting from the double-difference method. Due to the $IP$ removal, the butterfly patterns show more distinct features than the $Q^+$ and $U^+$ images and the recorded noise outside of the disk is reduced. 

An additional correction is made for the $IP$ introduced upstream of the HWP, following the method outlined in \citet{Canovas_2011} and \citet{de_Boer_2019}. The correction is performed for each HWP cycle to mitigate temporal differences in the $IP$ as a result of changing angles of reflection. As before, it is assumed that the stellar light is unpolarised and polarised signal near the star is ascribed to instrumental polarisation \citep{Quanz_2011}. The median $Q/I$ signal is computed over the same annulus $i$ from Eq. \ref{eq:X_oe} to obtain a scalar $c_Q$. To obtain $c_U$, we calculate the median $U/I$ signal over the same annulus. Per annulus, the $IP$-subtracted linear Stokes components are found by subtracting the product of these scalars and the respective $I_Q$ or $I_U$ image:
\begin{align}
    Q_\mathrm{IPS} &= Q - I_Q\cdot c_Q, \label{eq:Q_IPS} \\
    U_\mathrm{IPS} &= U - I_U\cdot c_U. \label{eq:U_IPS}
\end{align}
By using multiple user-specified annuli, the pipeline retrieves various $IP$-subtracted results. The third row of panels in Fig. \ref{fig:QU} displays the median $Q_\mathrm{IPS}$ and $U_\mathrm{IPS}$ images where the annulus was drawn between a radius of 3 and 6 pixels. As expected from the correction, the $Q_\mathrm{IPS}$ measurement shows a decreased signal near the star compared to the $Q$ image. 

In Fig. \ref{fig:QU}, the $U_\mathrm{IPS}$ signal is lower than $Q_\mathrm{IPS}$ as a result of crosstalk between the linear and circular Stokes components \citep{Witzel_2011}. If a disk is unmistakably
 detected and approximately axisymmetric, this reduced efficiency of the Stokes $U$ component relative to $Q$ can be estimated following the method outlined by \citet{Avenhaus_2014a}. In an annulus with disk signal, the number of pixels where $|Q_\mathrm{IPS}|>|U_\mathrm{IPS}|$ is expected to be equal to the number of pixels where $|U_\mathrm{IPS}|>|Q_\mathrm{IPS}|$. We can multiply the $U_\mathrm{IPS}$ image by a factor of $1/e_U$ so that the above assumption holds. The crosstalk-corrected components are then
\begin{align}
    Q_\mathrm{CTC} &= Q_\mathrm{IPS}, \\
    U_\mathrm{CTC} &= \frac{1}{e_U}\cdot U_\mathrm{IPS},
\end{align}
where we assume an efficiency of $100\%$ for Stokes $Q$. By modelling the NACO $IP$ with standard star observations, \citet{Millar_Blanchaer_2020} conclude that the Stokes $Q$ has an efficiency of $\sim$\,$90\%$. Since such a correction is not performed with PIPPIN, any quantitative polarimetry measurements on the reduced data products could be off by $\sim$\,$10\%$. The efficiency-correction should not be performed in instances with ambiguous signal and thus PIPPIN only makes the crosstalk-correction if requested.

Incomplete HWP cycles, with only measurements of $Q^\pm$ (or $U^\pm$), are removed. If only the Stokes $Q^+$ and $U^+$ (or only $Q^-$ and $U^-$) were recorded, PIPPIN will still be able to produce the final data products, but the double-difference method cannot be applied. At this point, the pipeline computes the median $Q$, $U$, $I_Q$, $I_U$, and $I$ over all observations. The final polarisation images ($PI$, $Q_\phi$, and $U_\phi$) are described below in terms of $Q$ and $U$, but we note that these data products are also calculated with $Q_\mathrm{IPS}$/$U_\mathrm{IPS}$ and $Q_\mathrm{CTC}$/$U_\mathrm{CTC}$, if possible. The total polarised intensity is calculated as\begin{align}
    PI = \sqrt{Q^2+U^2}.
\end{align}
This method of squaring $Q$ and $U$ can lead to the increase in noise in regions where the $Q$ or $U$ signal originating from the disk is low. A cleaner image can be found with the azimuthal Stokes parameters outlined in \citet{Monnier_2019} and \citet{de_Boer_2019}, analogous to \citet{Schmid_2006}, but with a flipped sign:
\begin{align}
    Q_\phi &= -Q\cos(2\phi) -U\sin(2\phi), \\
    U_\phi &= +Q\sin(2\phi) -U\cos(2\phi),
\end{align}
where $\phi$ is the azimuthal angle and is calculated for each pixel with\begin{align}
    \phi &= \arctan\left(\frac{\textit{y}-\textit{y}_\mathrm{star}}{\textit{x}_\mathrm{star}-\textit{x}}\right) + \phi_0, \label{eq:phi}
\end{align}
where ($\textit{x}_\mathrm{star}$, $\textit{y}_\mathrm{star}$) are the pixel-coordinates of the central star. If the disk has a low inclination and the scattered light emerges from single scattering events, the polarisation is oriented azimuthally with respect to the star. Consequently, the $Q_\phi$ image shows a positive signal as it measures polarisation angles of $\pm90^\circ$. Simultaneously, the $U_\phi$ image is expected to show a negligible signal as it measures polarisation angles of $\pm45^\circ$. However, a non-zero $U_\phi$ signal can occur if there is crosstalk between $Q$ and $U$, if the light is scattered multiple times \citep{Canovas_2015b}, if the disk has a high inclination, and if an inadequate correction retains stellar or instrumental polarisation \citep{Hunziker_2021}. If requested, PIPPIN can minimise the $U_\phi$ signal in the same annulus used for the crosstalk-correction by fitting for the azimuth angle offset $\phi_0$, similar to \citet{Avenhaus_2014a}. Otherwise, the offset angle $\phi_0$ is set to $0^\circ$.

The median total intensity $I$, polarised intensity $PI$, and azimuthal Stokes parameters $Q_\phi$ and $U_\phi$ with different levels of $IP$ correction are shown in Fig. \ref{fig:PI} for HD 135344B. Once PIPPIN has computed the final data products, these images are de-rotated using \texttt{scipy.ndimage.rotate}. Therefore, contrary to Fig. \ref{fig:QU}, the panels of Fig. \ref{fig:PI} have north pointing up and east to the left. It is apparent from the total and polarised intensity images that the PDI technique applies an extremely effective suppression of the stellar signal, thus revealing the circumstellar disk and its spiral arms. In this example, we observe the $U_\phi$ signal diminish as the $IP$ corrections are performed. Since HD 135344B is observed at a low inclination and axisymmetric to a first order, we employed crosstalk-correction and $U_\phi$ minimisation to produce the final Stokes images. For these Ks-band observations, we find a reduced efficiency of $e_U=0.65$, in agreement with \citet{Garufi_2013} who find an efficiency of $0.61$ \citep{Avenhaus_2014a}. Similarly, the more extensive $IP$ model presented by \citet{Millar_Blanchaer_2020} results in an efficiency of $e_U=0.7\pm0.02$ for Elia 2-25. Furthermore, we find an offset angle of $\phi_0=5.3^\circ$, while $3.7^\circ$ is derived in the previous analysis of these data \citep{Garufi_2013,Avenhaus_2014a}.

\subsubsection{Non-HWP and wire-grid observations} \label{sect:rotator_wiregrid}
Prior to August 8, 2003, NACO was not equipped with a HWP. Rather than rotating the HWP to modulate the direction of polarisation, observers would alter the PA by rotating the instrument on its rotator ring. PIPPIN automatically diagnoses whether the de-rotator flange of the telescope support structure was used \citep{Lenzen_2003}. For these data, the HWP angles $\theta=0$, $22.5$, $45$, and $67.5^\circ$ in Eqs. \ref{eq:Q+}, \ref{eq:Q-}, \ref{eq:U+}, and \ref{eq:U-} are replaced by the PAs of the instrument: $\theta_{PA}=0$, $45$, $90$, and $135^\circ$. The $Q^\pm$, $U^\pm$, $I_{Q^\pm}$, and $I_{U^\pm}$ images are also de-rotated to align the circumstellar structures before combining them with Eqs. \ref{eq:double_diff1}, \ref{eq:double_diff2}, \ref{eq:double_sum1}, and \ref{eq:double_sum2}. For the rotator observations, the $IP$ subtraction of Eqs. \ref{eq:Q_IPS} and \ref{eq:U_IPS} is also performed. 

In the early stages of its operation, NACO was equipped with wire grids to carry out polarimetric observations, rather than the Wollaston prism. In our cross-validation of the ESO archive, we found four potentially young sources that were observed in this manner: V1647 Ori, NX Pup, Mon R2 IRS 3, and R Mon. PIPPIN adopts the Pol\_00, Pol\_45, Pol\_90, and Pol\_135 wire grids as measurements of the Stokes $Q^+$, $U^-$, $Q^-$, and $U^+$ components, respectively. The linear Stokes components are propagated in the presence of the HWP. The only beam that is present in the images is fit with a single Moffat function. Since the wire-grid observations are not limited by the height of the polarimetric mask, their final data products have a much larger field of view than those obtained with the Wollaston prism.

\subsubsection{Supplemental data products}
Since the disk is illuminated by the star, the scattered light brightness decreases by the inverse of the squared distance to the host star. To better visualise structures at larger separations from the star, PIPPIN also produces images that are multiplied by the squared, de-projected radius. The disk PA, $PA_\mathrm{disk}$, is used to calculate the offsets along the major axis, $\Delta\textit{x}_\mathrm{disk}$, and minor axis, $\Delta\textit{y}_\mathrm{disk}$, with
\begin{align}
    \Delta\textit{x}_\mathrm{disk} &= \Delta(\mathrm{R.A.})\cdot\sin PA_\mathrm{disk} + \Delta(\mathrm{Dec.})\cdot\cos PA_\mathrm{disk}, \\
    \Delta\textit{y}_\mathrm{disk} &= \Delta(\mathrm{Dec.})\cdot\sin PA_\mathrm{disk} - \Delta(\mathrm{R.A.})\cdot\cos PA_\mathrm{disk},
\end{align}
where $\Delta(\mathrm{R.A.})$ and $\Delta(\mathrm{Dec.})$ are the right ascension and declination offsets with respect to the star. Subsequently, the de-projected radius $r$ is computed with
\begin{align}
    r &= \sqrt{\Delta\textit{x}_\mathrm{disk}^2 + \left(\frac{\Delta\textit{y}_\mathrm{disk}}{\cos i_\mathrm{disk}}\right)^2},
\end{align}
where $i_\mathrm{disk}$ is the disk inclination. As is shown in Table \ref{tab:PIPPIN_keywords}, the disk PA, $PA_\mathrm{disk}$ and inclination $i_\mathrm{disk}$ are specified in the configuration file for PIPPIN and are set to $0^\circ$ by default. In cases where the disk inclination and PAs are unknown, the default values ensure that the images are scaled by the projected separation from the host star.

The height of the final data products is limited to $\sim$\,$3.0\ \mathrm{arcsec}$ (using the S27 detector) due to the polarimetric mask. Observations where the PA was rotated, rather than the HWP, cover a larger area of the sky. Since the sky rotates while the polarimetric mask remains stationary, the effective field of view is increased. Figure \ref{fig:gallery_embedded_with_QU} depicts this increased sky coverage. An eight-pointed star emerges where at least one $Q$ and one $U$ component are covered and thus the polarised intensity can be computed within this shape. An inner octagon appears where every positive and negative Stokes component is observed. We note that the signal-to-noise decreases for areas outside of this octagon, due to the reduced number of observations. PIPPIN outputs the extended eight-pointed star images in addition to the data products resulting from the double-difference method, which are restricted to the inner octagon that has a complete coverage.

\section{Inspection of NACO data} \label{Section3}
\subsection{Identification of detections} \label{sect:cross_corr}
\begin{figure}[h!]
    \resizebox{\hsize}{!}{\includegraphics[width=17cm]{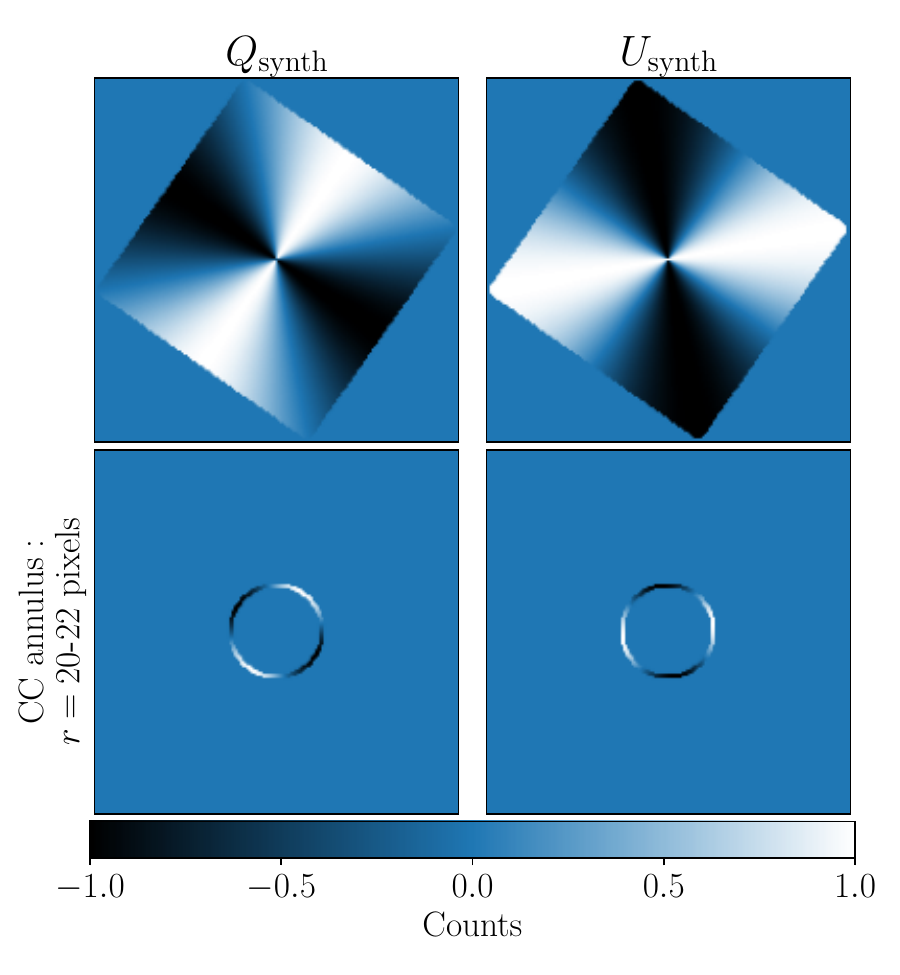}}
    \caption{Templates for observations of HD 135344B with a PA of $-35^\circ$. \textit{Top panels}: Complete $Q_\mathrm{synth}$ and $U_\mathrm{synth}$ templates. Values range from $-1$ to $+1,$ and pixels outside of the image are set to $0$. \textit{Bottom panels}: Example annuli used in computing the cross-correlation coefficients.}
    \label{fig:cross_corr_example}
\end{figure}

\begin{figure*}[h!]
    \centering
    \includegraphics[width=17cm]{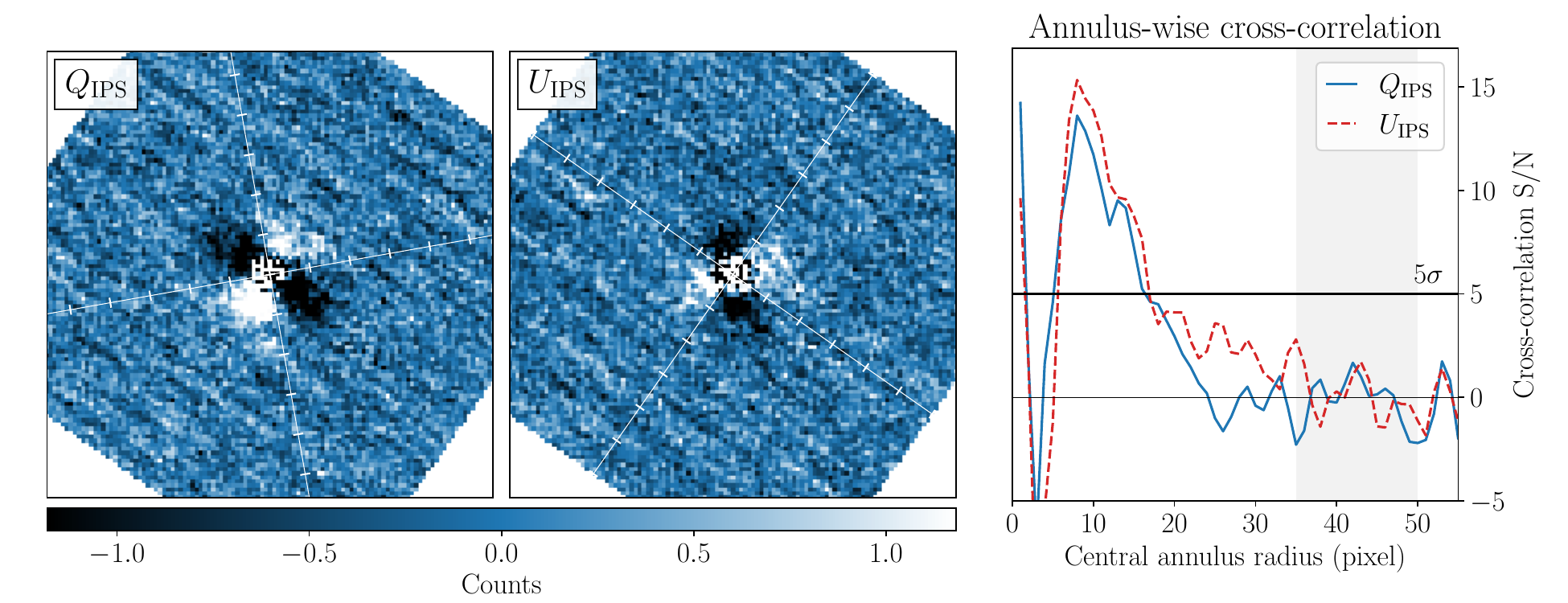}
    \caption{Detection analysis of HD 135344B observed in the narrowband NB\_1.64. \textit{Left panels}: $Q_\mathrm{IPS}$ and $U_\mathrm{IPS}$ images divided by the white lines into the four quadrants of the expected butterfly pattern. \textit{Right panel}: Annulus-wise CCFs, with the $\mathrm{S/N}$ shown against the annulus radius in pixels. The results for the $Q_\mathrm{IPS}$ and $U_\mathrm{IPS}$ images are plotted in blue and red, respectively. The shaded region specifies the coefficients used in normalising and converting the CCF into a $\mathrm{S/N}$ function. The $5\sigma$ detection limit is indicated with a horizontal line.}
    \label{fig:HD_135344B_detection}
\end{figure*}

The PIPPIN pipeline described above was used to reduce all observations listed in Table \ref{tab:systems}. The table lists multi-epoch, multi-wavelength observations as well as different exposure times and whether the wire grids were used or the Wollaston prism, with the HWP or PA. For each set of observations, we indicate the (non-)detection of circumstellar material in the final data products. The detection significance of polarised signal was assessed via a template-matching method, akin to cross-correlation, applied to the Stokes Q and U images. In the case of a detection, we expect that the signal is present in multiple, adjacent pixels and forms a specific butterfly pattern. Synthetic $Q_\mathrm{synth}$ and $U_\mathrm{synth}$ templates of the expected butterfly patterns were constructed with\begin{align}
    Q_\mathrm{synth} &= -\cos\big(2(\phi-PA)\big), \\
    U_\mathrm{synth} &= -\sin\big(2(\phi-PA)\big),
\end{align}
where $\phi$ is the azimuthal angle calculated with Eq. \ref{eq:phi} and $PA$ is the position angle of the observation, which is subtracted since PIPPIN de-rotates the final data products, including the $Q_\mathrm{IPS}$ and $U_\mathrm{IPS}$ images. Subsequently, the $Q_\mathrm{synth}$ and $U_\mathrm{synth}$ templates were divided into multiple annuli with increasing radius and width of $2\ \mathrm{pixels}$, roughly corresponding to one resolution element in the H  ($42\ \mathrm{mas}$) and Ks band ($56\ \mathrm{mas}$) at a pixel scale of $27\ \mathrm{mas\ pixel^{-1}}$. Figure \ref{fig:cross_corr_example} shows an example of the $Q_\mathrm{synth}$ and $U_\mathrm{synth}$ templates and a single annulus for a PA of $-35^\circ$, corresponding to the observations of HD 135344B. The values in the templates range from $-1$ to $+1$ and pixels outside of the annulus are set to $0$, thus ensuring that they do not contribute when calculating the cross-correlation coefficient. In annulus $i$, a cross-correlation coefficient is calculated for the $Q$ and $U$ signals:
\begin{align}
    CC_{Q,i} &= \sum_{\mathrm{pixels}} Q_{\mathrm{IPS},i}\cdot Q_{\mathrm{synth},i}, \label{eq:CC_Q} \\
    CC_{U,i} &= \sum_{\mathrm{pixels}} U_{\mathrm{IPS},i}\cdot U_{\mathrm{synth},i}, \label{eq:CC_U}
\end{align}
where the sum is performed over every pixel within annulus $i$. In this manner, a positive pixel increases the coefficient if the respective quadrant expects a positive signal. A negative signal in the negative quadrants of the template also contributes, whereas a discrepant signal reduces the cross-correlation coefficient. A cross-correlation function (CCF) is constructed by computing a coefficient for each annulus. For the narrowband NB\_1.64 observations of HD 135344B, the rightmost panel of Fig. \ref{fig:HD_135344B_detection} displays the CCFs for the $Q_\mathrm{IPS}$ and $U_\mathrm{IPS}$ components in blue and red, respectively. The CCF was converted into a signal-to-noise ($\mathrm{S/N}$) function by subtracting the mean coefficient between $35$ and $50$ pixels, and subsequently dividing by the standard deviation of coefficients within that same range, indicated by the grey shaded region in the right panel. The annulus-wise CCFs peak at a radius of 8 pixels with signal-to-noises of $\mathrm{S/N}\sim$\,$13$ and $\sim$\,$15$, respectively for $Q_\mathrm{IPS}$ and $U_\mathrm{IPS}$. These maxima surpass our $5\sigma$ detection threshold, thereby identifying this observation as a detection. Although the template-matching method generally worked well, it failed to flag two observations of HR 4796 as detections, despite the polarised signal evident from a visual inspection. These non-detections can be ascribed to the high inclination and narrow features of HR 4796, while the outlined template-matching analysis works optimally for face-on disks. 

In this reduction of the NACO data, many of the non-detections are likely the result of small or faint disks, or the absence of polarised light. Notably, IM Lup, GQ Lup, and EX Lup do not show polarised light in the NACO data, despite their prominent detections with SPHERE/IRDIS \citep{Avenhaus_2018,van_Holstein_2021,Rigliaco_2020}. The data of IM Lup and GQ Lup were not previously published, whereas \citet{Kospal_2011} also report a non-detection of polarised light in the EX Lup observations.

\subsection{Analysis of detected polarised light}
\begin{figure*}[h]
    \centering
    \includegraphics[width=17cm]{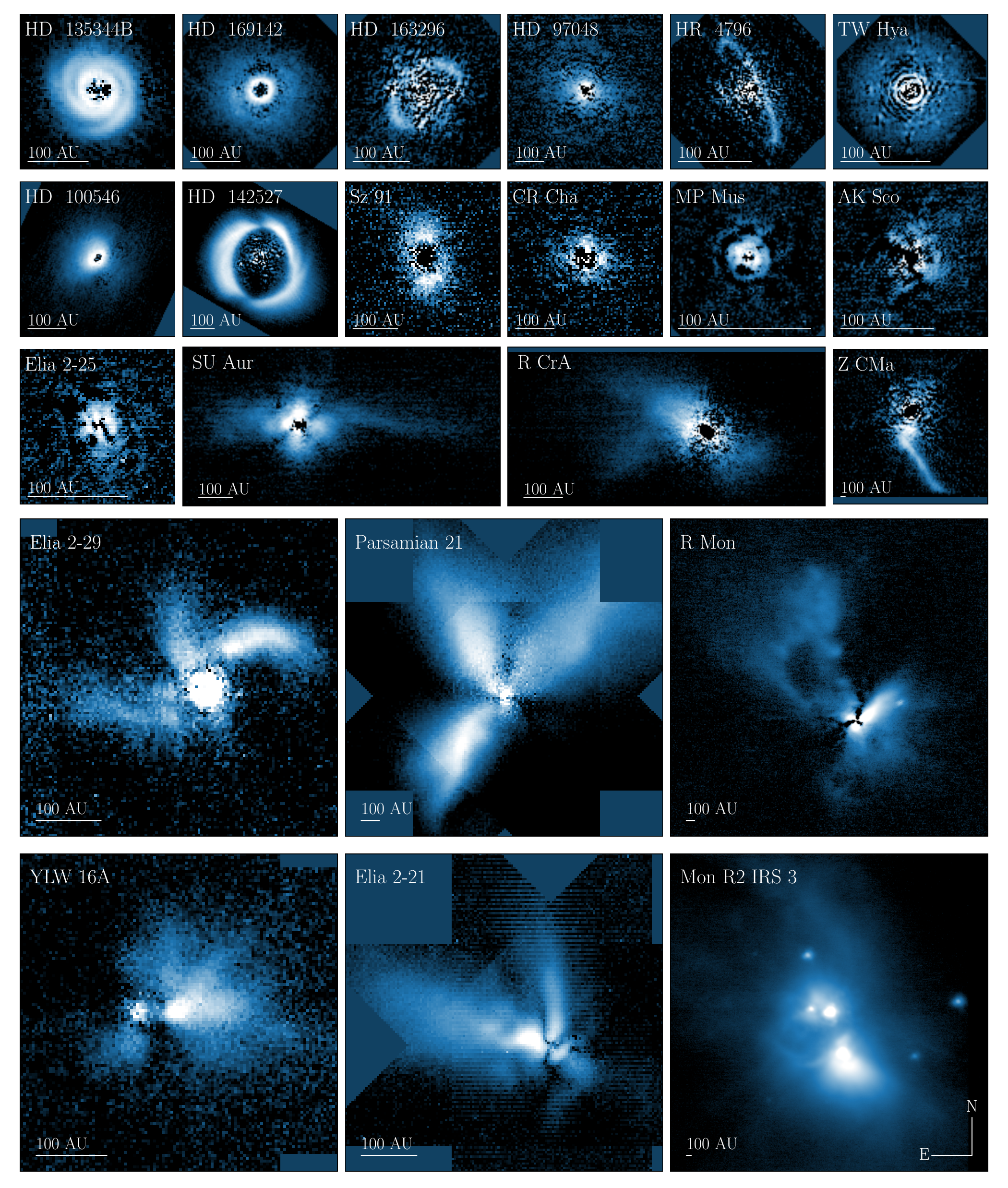}
    \caption{Gallery of young systems detected with NACO and reduced with PIPPIN. Each panel shows the polarised light on a logarithmic scale ranging between different values to highlight sub-structures. The highest degree of $IP$ correction is used where possible. Scale bars in the lower-left corners of each panel indicate $100\ \mathrm{AU}$ at each object's distance. HD 169142, R CrA, and Parsamian 21 are shown in the H band, MP Mus is shown in the IB\_2.06 filter, and the other panels use Ks-band observations. Mon R2 IRS 3 shows the median $I_Q$ image because the Stokes $U$ component was not observed. The images of YLW 16A and Elia 2-21 present the first polarised light detections in the NACO observations.}
    \label{fig:gallery}
\end{figure*}

As demonstrated in Table \ref{tab:systems}, in 22 out of the 57 observed systems, we find at least one set of observations with polarised signal originating from circumstellar material. Figure \ref{fig:gallery} presents a gallery of these detections and highlights a diverse collection of morphologies. As mentioned in Sect. \ref{Section2}, HD 135344B shows distinct spiral arms in its circumstellar disk while HD 142527 has spiral features in its eastern and western lobes. Furthermore, we detect rings in a large number of disks, including HD 169142, HD 163296, HD 97048, HR 4796, TW Hya, HD 142527, and Sz 91. HR 4796 is the only debris disk in our sample and its $Q_\phi$ image shows a narrow ring. The disks around HD 163296, HD 97048, and HD 100546 are offset from the central star, suggesting that the scattering surface is above the disk midplane as confirmed by \citet{Monnier_2017}, \citet{Ginski_2016}, and \citet{Sissa_2018}. The highly extended disk of HD 142527 shows a large inner cavity that is possibly cleared out by an inner companion \citep{Biller_2012,Close_2014,Lacour_2016,Claudi_2019}, undetected in polarised light. Moreover, we find narrow shadow lanes imprinted on the disks of HD 142527 and SU Aur, similar to \citet{Avenhaus_2017} and \citet{Ginski_2021}. In these cases, misaligned inner disks prevent the stellar light from reaching certain areas of the outer disk. CR Cha, MP Mus, AK Sco, and Elia 2-25 show negligible structure due to their small sizes, but a significant butterfly pattern was detected, leading to their inclusion in Fig. \ref{fig:gallery}. As a possible consequence of their small sizes, the polarimetric NACO data of CR Cha and MP Mus were previously unpublished.

Figure \ref{fig:gallery} displays a number of objects with extended features that appear inconsistent with circumstellar disks. SU Aur shows tail-like features, where \citet{Ginski_2021} discover an inflow of material onto the disk by combining polarimetric SPHERE observations with ALMA line observations. The features of R CrA resemble the non-polarimetric SPHERE observations presented by \citet{Mesa_2019} with scattered light towards the north-east, south-east, and south-west of the primary star. Although a brightness asymmetry is observed towards the east in the NACO $Q_\phi$ image, the companion inferred by \citet{Mesa_2019} is not detected. To our knowledge, the detection of polarised light in the NACO observations of R CrA went unpublished before this work. Recently, \citet{Dong_2022} reported that the binary star Z CMa experienced a close encounter with a nearby star (masked in the NACO observation), thereby ejecting the streamer structure that we also observe in the $Q_\phi$ image. As YLW 16A, Elia 2-29, Elia 2-21, and Parsamian 21 were observed with the rotator flange, Fig. \ref{fig:gallery} presents the extended data products described in Sect. \ref{Section2}. The polarised light of Elia 2-29 reveals three arcs to the east, north and north-west of the central star. The northern and north-western arcs have curved shapes that are reminiscent of spiral-like features \citep{Huelamo_2007}. YLW 16A shows asymmetric polarised signal to the west and north-west of the binary components \citep{Plavchan_2013} that are still visible as intensity maxima. Parsamian 21 and Elia 2-21 appear to show bipolar outflows along the NW-SE and NE-SW axes, respectively. Both nebulae are distinctly asymmetric with the northern and north-eastern segments showing the largest emission surfaces, respectively for Parsamian 21 and Elia 2-21. At large separations, along the PAs of the $Q^\pm$ and $U^\pm$ measurements, one of the linear Stokes components dominates over the other. Hence, the majority of the signal in $PI$ can be represented by the absolute values $|Q^\pm|$ or $|U^\pm|$, which is shown with a grey colour map in Fig. \ref{fig:gallery_embedded_with_QU}. The northern lobe of Parsamian 21 and the northern and eastern arms of Elia 2-21 are traced about $\sim$\,$2\ \mathrm{arcsec}$ further. To our knowledge, this work is the first publication of the polarimetric NACO observations of Elia 2-21 and YLW 16A. The polarised light of the reflection nebula NGC 2261, illuminated by R Mon, demonstrates distinct emission from the extended north-eastern and south-western components. The Stokes U component of the infrared source Mon R2 IRS 3, part of the Monoceros R2 molecular cloud complex, was not observed. Hence, Fig. \ref{fig:gallery} presents the median $I_Q$, which also includes unpolarised (scattered) light. Despite this, filamentary structure is unambiguously detected. As discussed in Sect. \ref{sect:rotator_wiregrid}, the two detections utilising polarimetric wire grids, R Mon and Mon R2 IRS 3, have much larger image sizes than the regular data products (i.e. compared to the upper rows of panels in Fig. \ref{fig:gallery}).

\subsection{Disk classification and brightness}
Circumstellar polarised light is detected in nine out of the 14 Herbig stars in our sample. Grouping the Orion variables together with T Tauri stars, we detect polarised signal in eight out of 28 systems. For the YSOs, five of the 14 sources are flagged as detections with the template-matching analysis. The only high-proper motion star in our sample, HR 4796, also constitutes the only detection of a debris disk. However, our selection of young stars based on the SIMBAD object type could have missed some of the older Class III disks. Since NACO frequently observed disks known to be extended and thus potentially observable in scattered light, the gathered sample is certainly not unbiased. For that reason, a statistical analysis of the disk occurrence per object type is somewhat arbitrary.

\begin{figure*}[h!]
    \centering
    \includegraphics[width=17cm]{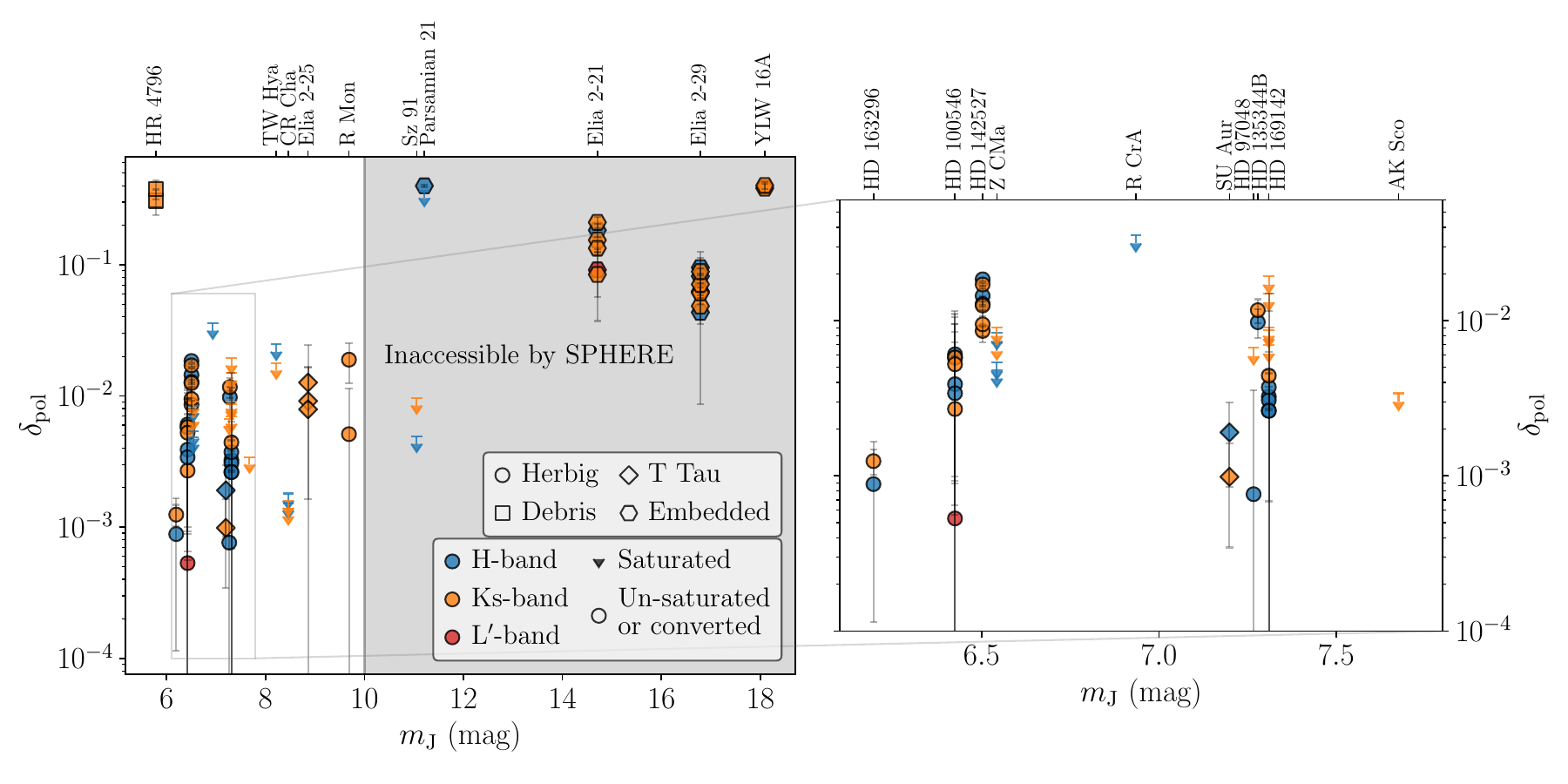}
    \caption{Polarised-to-stellar light contrast, $\delta_\mathrm{pol}$, plotted against the apparent J-band magnitude. The \textit{right panel} shows a zoomed-in view of the bright $m_\mathrm{J}$. The object names are listed along the top axes. The marker colours and symbols specify the observing filter and object type, respectively. Upper limits are shown when the stellar PSF was determined to be saturated. The error bars show the $3\sigma$ uncertainties. The grey shaded region shows the approximate magnitudes ($m_\mathrm{J}\gtrsim10$) inaccessible by the SPHERE AO system.}
    \label{fig:disk_contrast_vs_m_J}
\end{figure*}

Here we examine the disk brightness of the sample of circumstellar disks detected by NACO. Since the disk inclination, disk extent, stellar brightness, and the distance to the source affect the total disk brightness, we made use of the polarised-to-stellar light contrast $\delta_\mathrm{pol}$ \citep{Garufi_2017,Garufi_2022b,Benisty_2022}. The polarised flux per unit area, $F_\mathrm{pol}$, was multiplied by the squared separation $4\pi r^2$ to account for the reduced stellar illumination. Subsequently, we normalised it by the stellar flux $F_*$ and averaged radially along the disk's major axis. Thus, the polarised-to-stellar light contrast was computed as\begin{align}
    \delta_\mathrm{pol} &= \frac{1}{r_\mathrm{out}-r_\mathrm{in}}\cdot \int_{r_\mathrm{in}}^{r_\mathrm{out}} \frac{F_\mathrm{pol}(r)}{F_*}\cdot 4\pi r^2 dr, \label{eq:delta_pol}
\end{align}
where $r_\mathrm{in}$ and $r_\mathrm{out}$ are the inner- and outermost radii, respectively. This measurement expresses the fraction of stellar photons that became polarised as a result of scattering by the resolved disk. The contrast $\delta_\mathrm{pol}$ is set by the geometry and composition of the circumstellar disk and is sometimes referred to as the geometrical albedo. Following the method outlined in Appendix B of \citet{Garufi_2017}, we performed a 2-pixel-wide cut (one resolution element) of the $Q_\phi$ image along the major axis of the disk. The photons measured along the major axis are scattered with angles close to $90^\circ$. This cut reduces the impact of the disk inclination on its brightness due to any asymmetry in the scattering phase function. The PA of the major axis varies for the sources observed by NACO, but was set to $0^\circ$ when this axis could not be estimated. These ambiguous sources (e.g. CR Cha) were roughly azimuthally symmetric and therefore did not significantly affect the derived contrast. The inner- and outermost radii ($r_\mathrm{in}$, $r_\mathrm{out}$) were determined by eye for each system detected in scattered light. The disk inclination and scale height were not taken into account when re-scaling the polarised flux by the projected separation. Similar to \citet{Garufi_2017}, we calculated the primary error of $\delta_\mathrm{pol}$ by deriving the standard deviation in a resolution element of the $Q_\phi$ image. Subsequently, this noise estimate for each pixel was propagated through Eq. \ref{eq:delta_pol} to find $\sigma_{\delta_\mathrm{pol}}$.

Fig. \ref{fig:disk_contrast_vs_m_J} displays the polarised-to-stellar light contrast $\delta_\mathrm{pol}$ against the apparent J-band magnitude $m_\mathrm{J}$, measured as part of the 2 Micron All Sky Survey (2MASS; \citealt{Cutri_2003}). The source names are shown along the top axes. Blue, orange, and red markers indicate whether the observation was performed in the H, Ks, or L' band. Diamonds (T Tau), circles (Herbig), and squares (debris) mark the object types. We note that HD 135344B is shown as a Herbig star (circle), in line with \citet{Garufi_2014}, but contrary to the SIMBAD object type in Table \ref{tab:systems}. Similarly, Parsamian 21 is depicted as an embedded YSO despite its Orion variable type reported in Table \ref{tab:systems}. The $\delta_\mathrm{pol}$ values of saturated PSFs are presented as upper limits in Fig. \ref{fig:disk_contrast_vs_m_J} (triangles; 99.75-th percentile) because the stellar flux $F_*$ is underestimated. In some instances, the source was also observed with narrowband filters where the stellar PSF was not saturated due to the smaller filter width. Hence, we could estimate the broadband flux $F_*^\mathrm{BB}$, using the narrowband flux $F_*^\mathrm{NB}$, if the two filters had overlapping wavelength ranges. We calculated the stellar flux as\begin{align}
    F_*^\mathrm{BB} &= F_{*}^\mathrm{NB} \cdot \frac{t_\mathrm{exp}^\mathrm{BB}}{t_\mathrm{exp}^\mathrm{NB}} \cdot \frac{\int T^\mathrm{BB}(\lambda)\ \mathrm{d}\lambda}{\int T^\mathrm{NB}(\lambda)\ \mathrm{d}\lambda},
\end{align}
where $t_\mathrm{exp}^\mathrm{BB}$ and $t_\mathrm{exp}^\mathrm{NB}$ are the exposure times of the broadband and narrowband observations, respectively. We integrated over the corresponding transmission curves $T(\lambda)$ to estimate how many photons should be detected for each photon in the narrowband filter. For the H band, we used the NB\_1.64 and NB\_1.75 narrowband filters. For the Ks band, NB\_2.17, IB\_2.18, NB\_2.12, NB\_2.15, and IB\_2.21 were employed to compute the broadband flux and the NB\_3.74 filter was used for saturated L'-band observations. 

Since NACO was equipped with a NIR WFS, it could observe sources down to $\mathrm{K}\approx 14\ \mathrm{mag}$ \citep{Rousset_2003}. For comparison, SPHERE's optical WFS has a magnitude limit of $\mathrm{R}\approx14\ \mathrm{mag}$ \citep{Beuzit_2019}, GPI has a limit of $\mathrm{I}\approx10\ \mathrm{mag}$ \citep{Macintosh_2014}, and SCExAO/CHARIS on the Subaru telescope is limited by $\mathrm{R}\approx13\ \mathrm{mag}$\footnote{\url{https://www.naoj.org/Projects/SCEXAO/scexaoWEB/010usingSCExAO.web/010currentcap.web/020wavefrontcorrection.web/indexm.html}}. For that reason, NACO could achieve a unique insight into embedded protostars, despite their faint optical magnitudes. The grey shaded region in Fig. \ref{fig:disk_contrast_vs_m_J} roughly indicates the sources that are inaccessible by the optical WFS installed on SPHERE. Since the estimated J-band magnitude limit of SPHERE depends on the spectral type of the observed source, the limit of $m_\mathrm{J}\sim$\,$10\ \mathrm{mag}$ should be viewed as a crude assessment. From Fig. \ref{fig:disk_contrast_vs_m_J} we find that the four embedded protostars Parsamian 21, Elia 2-21, Elia 2-29, and YLW 16A, in addition to the low-mass star Sz 91, are likely not observable with modern PDI instruments.

\begin{figure*}[h!]
    \centering
    \includegraphics[width=16cm]{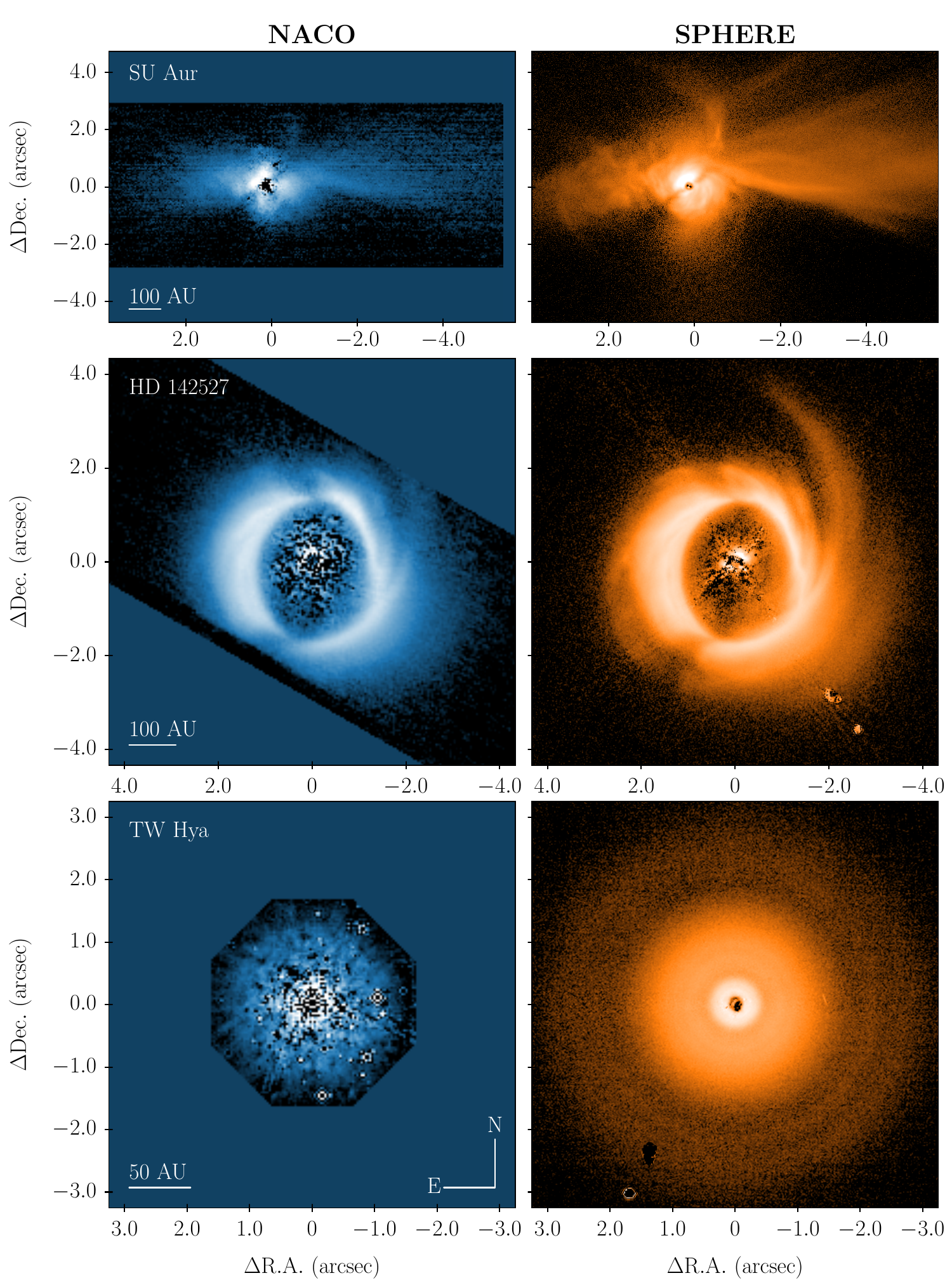}
    \caption{Comparison between PIPPIN-reduced NACO $Q_\phi$ observations (\textit{left panels}) and the more recent SPHERE data (\textit{right panels}). \textit{From top to bottom}: SU Aur, HD 142527, and TW Hya observed in the H band with both instruments. The SPHERE observations were previously published by \citet{Ginski_2021}, \citet{Hunziker_2021}, and \citet{van_Boekel_2017} for SU Aur, HD 142527, and TW Hya, respectively.}
    \label{fig:SPHERE_comparison}
\end{figure*}

The polarised-to-stellar light contrasts $\delta_\mathrm{pol}$ are listed in Table \ref{tab:systems}. We note that the $\delta_\mathrm{pol}$ values are possibly underestimated by $\sim$\,$10\%$ (see Sect. \ref{sect:PDI}) due to the absence of a correction for the reduced $Q$ efficiency resulting from crosstalk. For the sources with available mass estimates (also included in Table \ref{tab:systems}), we fail to detect a trend between $\delta_\mathrm{pol}$ and the stellar mass $M_*$. Since the disk's dust mass is related to the stellar mass (e.g. \citealt{Pascucci_2016}), the absence of a distinct trend reveals that the disk brightness in polarised light is not strongly correlated with the abundance of dust in the system. Instead, the scattered light brightness is affected by the amount of light that is intercepted. The geometry of the system primarily influences the polarised-to-stellar light contrast $\delta_\mathrm{pol}$, with the dust composition acting as a secondary effect. Similarly, \citet{Garufi_2022b} find no correlation between the disk brightness in scattered light and dust mass estimated from the $1.3\ \mathrm{mm}$ flux. We also find no apparent distinction in disk brightnesses between the T Tauri stars and Herbig stars. Comparing the obtained $\delta_\mathrm{pol}$ results with those presented in Fig. 3 of \citet{Garufi_2022b} and Table A.1 of \citet{Garufi_2017}, we find good agreement for HD 163296, HD 100546, HD 142527, HD 97048, HD 135344B, HD 169142, AK Sco, TW Hya, and CR Cha. For the extended systems SU Aur, R CrA, and Z CMa, we assessed the polarised-to-stellar light contrast ratio of a potential disk component at close separations, meaning that $r_\mathrm{out}$ was limited to $25$, $25,$ and $53$ pixels, respectively. We find $\delta_\mathrm{pol}\sim$\,$1.5\cdot10^{-3}$ for SU Aur, $\lesssim3\cdot10^{-2}$ for R CrA, and $\lesssim6\cdot10^{-3}$ for Z CMa. The brightest disk is found around HR 4796 with $\delta_\mathrm{pol}\sim$\,$0.3-0.4$. This finding is somewhat surprising, given that it is a flat debris disk and therefore should not intercept much stellar light. However, the high brightness is also reported in previous works where it is argued that the scattering phase functions are consistent with large ($\sim$\,$20\ \mathrm{\mu m}$) aggregate dust particles composed of small monomers \citep{Milli_2017,Milli_2019}. For Sz 91, the lowest-mass star ($M=0.58\pm0.07 M_\odot$; \citealt{Mauco_2020}) where polarised light is detected, we determine upper limits of $\delta_\mathrm{pol}\lesssim8\cdot10^{-2}$ and $\lesssim4\cdot10^{-2}$ in the Ks and H band, respectively. The estimated contrasts of multi-wavelength observations do not show any clear discrepancies between H- and Ks-band observations, owing to relatively large uncertainties. Hence, it is difficult to draw any conclusions about the dust composition by evaluating the disk colour.

\section{Discussion} \label{Section5}
The morphologies shown in Fig. \ref{fig:gallery} are almost identical to the polarised intensity images presented in previous publications of these data (see Table \ref{tab:systems} for references). The data reduction performed by PIPPIN therefore appears to reproduce the results obtained by the custom pipelines in other works. 

To study the performance between different instruments, Fig. \ref{fig:SPHERE_comparison} presents a comparison between the NACO and modern SPHERE observations of SU Aur (programme ID: 1104.C-0415(E), PI: Ginski), HD 142527 (programme ID: 099.C-0601(A), PI: Avenhaus), and TW Hya (programme ID: 095.C-0273(D), PI: Beuzit). In this comparison, we find the results of the different instrument characteristics. For instance, the NACO observations of TW Hya were made under better seeing conditions ($\sim$\,$0.5\ \mathrm{arcsec}$) than those made by SPHERE ($\sim$\,$0.7\ \mathrm{arcsec}$), but we find that the NACO polarised signal displays residual speckles over the circumstellar disk. The SPHERE $Q_\phi$ image does not show similar artefacts, likely due to the superior AO instrument. As part of the NACO instrument, NAOS had fewer actuators ($185$ active actuators for NAOS against $1377$ for SAXO; \citealt{Blanco_2022}) shaping the deformable mirror and its optical WFS operated at a lower frequency ($1200\ \mathrm{Hz}$ versus $444\ \mathrm{Hz}$; \citealt{Fusco_2006,Rousset_2003}), thus resulting in typical H-band Strehl ratios of $\sim$\,$10$ -- $35\%$ as opposed to $\sim$\,$60$ -- $80\%$ for SPHERE observations \citep{Fusco_2014,van_Boekel_2017}. Furthermore, the SPHERE NIR camera (IRDIS) has a pixel scale of $\sim$\,$12\ \mathrm{mas\ pixel^{-1}}$ \citep{Maire_2018} while the most-used S27 detector on CONICA had $\sim$\,$27\ \mathrm{mas\ pixel^{-1}}$. The NACO instrument was not equipped with a coronagraph in its polarimetric mode and thus short exposure times were utilised to avoid saturation by the central star. Each of the NACO observations presented in Fig. \ref{fig:SPHERE_comparison} employed considerably shorter single-frame integration times than the respective SPHERE observations (SU Aur: $0.35$ versus $32\ \mathrm{s}$, HD 142527: $0.3454$ versus $16\ \mathrm{s}$, TW Hya: $5$ versus $16\ \mathrm{s}$), thereby inevitably reducing the signal-to-noise. \citet{Ginski_2021} trace the extended western structure of SU Aur out to $\sim$\,$6\ \mathrm{arcsec}$, whereas the NACO data only confidently show signal out to $\sim$\,$4\ \mathrm{arcsec}$. Moreover, the filamentary structure observed in the tails and disk of SU Aur \citep{Ginski_2021} are not resolved in the NACO data due to the reduced signal-to-noise. Lastly, the polarimetric mask of NACO limits the vertical extent of the final data products to $\sim$\,$3\ \mathrm{arcsec}$. Hence, the north-western spiral structure of HD 142527 is eventually obscured in the NACO data.

\section{Conclusions} \label{Section6}
We have presented a complete catalogue of polarimetric NACO images for YSOs, reduced in a homogeneous manner with a new pipeline that employs the PDI technique. Via a cross-examination with the object types reported on SIMBAD, 57 targets were identified as potentially young systems with polarimetric NACO observations. As a result of multi-epoch and multi-wavelength observations, a total of 243 datasets were reduced with the publicly available PIPPIN pipeline\footnote{\url{https://pippin-naco.readthedocs.io}}. PIPPIN can handle observations made with NACO's HWP as well as its de-rotator. In addition to the Wollaston prism, observations measured with wire grids can be reduced too. Various levels of corrections for instrumental polarisation are performed, depending on the type of observation.

The reduced data products were analysed with a template-matching method to evaluate the detection significance. This technique exploits the butterfly pattern in the Stokes $Q$ and $U$ images that should be present in the case of significant polarised light. We find that 22 out of the 57 observed systems revealed polarised light in at least one observation. These detections revealed a wide diversity of sub-structures, including rings, gaps, spirals, shadows, and in- or outflowing matter. Since NACO was equipped with a NIR WFS, unique polarimetric observations of embedded YSOs were made. To our knowledge, this is the first work to publish the reduced data products of the Class I protostars Elia 2-21 and YLW 16A. PIPPIN also revealed detections of polarised light in the L' band for HD 100546 \citep{Avenhaus_2014b} and Elia 2-21. This long-wavelength filter ($3.8\ \mathrm{\mu m}$) is not available on current, state-of-the-art PDI instruments such as SPHERE/IRDIS, SCExAO/CHARIS, or GPI. 

Alongside this article, we publish an archive of the reduced data products generated with PIPPIN on Zenodo\footnote{\url{https://doi.org/10.5281/zenodo.8348803}}. As these observations were made in the past two decades, their combination with modern scattered light observations can be used to identify temporal changes in the sub-structures of planet-forming disks. In turn, such morphological changes can be used to infer the presence of a perturbing companion \citep{Ren_2020}. Recent studies of the NACO data of HD 97048 and SU Aur \citep{Ginski_2016,Ginski_2021} have led to the discovery of previously unidentified features. With this work, we hope to have outlined the utility of NACO observations reduced with the PDI technique.

\begin{acknowledgements}
S.d.R. acknowledges funding from the European Research Council (ERC) under the European Union’s Horizon 2020 research and innovation program under grant agreement No. 694513. 
C.C. acknowledges support by ANID BASAL project FB210003 and ANID, -- Millennium Science Initiative Program -- NCN19\_171. 
N.H. is funded by Spanish MCIN/AEI/10.13039/501100011033 grant PID2019-107061GB-C61.
S.P. acknowledges support from FONDECYT grant 1231663 and funding from ANID -- Millennium Science Initiative Program -- Center Code NCN2021\_080. 
MRS was supported from FONDECYT (grant number 1221059) and ANID, – Millennium Science Initiative Program – NCN19\_171. 
This work was also supported by the NKFIH excellence grant TKP2021-NKTA-64.
The authors thank all PIs of proposals whose observations are used in this work. This includes Ageorges, Apai, Christiaens, de Kok, Feldt, Hardy, Jeffers, Milli, Mouillet, Murakawa, Quanz, Schuetz, and Yudin.
This research has made use of the SIMBAD database, operated at CDS, Strasbourg, France. SPHERE is an instrument designed and built by a consortium consisting of IPAG (Grenoble, France), MPIA (Heidelberg, Germany), LAM (Marseille, France), LESIA (Paris, France), Laboratoire Lagrange (Nice, France), INAF – Osservatorio di Padova (Italy), Observatoire de Gen\`eve (Switzerland), ETH Z\"urich (Switzerland), NOVA (Netherlands), ONERA (France) and ASTRON (Netherlands) in collaboration with ESO. SPHERE was funded by ESO, with additional contributions from CNRS (France), MPIA (Germany), INAF (Italy), FINES (Switzerland) and NOVA (Netherlands). SPHERE also received funding from the European Commission Sixth and Seventh Framework Programmes as part of the Optical Infrared Coordination Network for Astronomy (OPTICON) under grant number RII3-Ct-2004-001566 for FP6 (2004–2008), grant number 226604 for FP7 (2009–2012) and grant number 312430 for FP7 (2013–2016). This research was performed with the \textit{Python} programming language. In particular, the \textit{SciPy} \citep{Virtanen_2020}, \textit{NumPy} \citep{Harris_2020}, \textit{Matplotlib} \citep{Hunter_2007}, and \textit{astropy} \citep{Astropy_2013,Astropy_2018} packages were utilised. 
\end{acknowledgements}

\bibliographystyle{aa}
\bibliography{References}

\clearpage
\onecolumn

\begin{appendix}

\begin{landscape}
\section{Reduced systems and observations} \label{app:table}
{\setlength{\tabcolsep}{3.5pt}
{\scriptsize
\begin{longtable}{lllll|lllllll|rr}
\caption{Potential young systems observed by NACO in its polarimetric mode, sorted by right ascension. \label{tab:systems}} \\

\hline\hline
\textbf{2MASS ID} & \textbf{Name} & \textbf{\begin{tabular}{@{}l}SIMBAD \\Object Type\end{tabular}} & \textbf{\begin{tabular}{@{}l}Spectral \\Type\end{tabular}} & $\mathbf{M\ (M_\odot)}$ & \textbf{Detection} & \textbf{\begin{tabular}{@{}l}HWP / \\PA / WG \end{tabular}} & \textbf{Obs. Date} & \textbf{Filter} & \textbf{Exp. Time (s)} & $\mathbf{N_\mathrm{\mathbf{obs}}}$ & $\mathbf{\delta_\mathrm{\mathbf{pol}}}$ & \textbf{Publication} & \textbf{programme ID} \\
\hline\hline
\endfirsthead

\caption{Continued.} \\

\hline\hline
\textbf{2MASS ID} & \textbf{Name} & \textbf{\begin{tabular}{@{}l}SIMBAD \\Object Type\end{tabular}} & \textbf{\begin{tabular}{@{}l}Spectral \\Type\end{tabular}} & $\mathbf{M\ (M_\odot)}$ & \textbf{Detection} & \textbf{\begin{tabular}{@{}l}HWP / \\PA / WG \end{tabular}} & \textbf{Obs. Date} & \textbf{Filter} & \textbf{Exp. Time (s)} & $\mathbf{N_\mathrm{\mathbf{obs}}}$ & $\mathbf{\delta_\mathrm{\mathbf{pol}}}$ & \textbf{Publication} & \textbf{programme ID} \\
\hline\hline
\endhead

\hline
\endfoot

J04292971+2616532 & FW Tau & Orion Var. & M5.8$^{(1)}$ &  & No & HWP & 2016-10-13 & Ks & 10 & 96 & - & - & 097.C-0644(A) \\
 &  &  &  &  & - & HWP & 2016-10-13 & Ks & 30 & 3 & - & - & 097.C-0644(A) \\
 &  &  &  &  & No & HWP & 2017-10-12 & Ks & 5 & 8 & - & - & 0100.C-0492(A) \\
 &  &  &  &  & No & HWP & 2017-10-12 & Ks & 20 & 8 & - & - & 0100.C-0492(A) \\
 &  &  &  &  & No & HWP & 2017-10-12 & Ks & 50 & 36 & - & - & 0100.C-0492(A) \\ \hline
J04294155+2632582 & DH Tau & Orion Var. & M2.3$^{(1)}$ &  & No & HWP & 2017-10-13 & Ks & 3 & 8 & - & - & 0100.C-0492(B) \\
 &  &  &  &  & No & HWP & 2017-10-13 & Ks & 55 & 36 & - & - & 0100.C-0492(B) \\ \hline
J04294247+2632493 & DI Tau & Orion Var. & M0.7$^{(1)}$ &  & No & HWP & 2017-10-13 & Ks & 10 & 8 & - & - & 0100.C-0492(B) \\ \hline
J04555938+3034015 & SU Aur & Orion Var. & G4$^{(1)}$ & $2.0\pm0.2^{(2)}$ & Yes & HWP & 2011-10-31 & H & 0.35 & 99 & $0.99\pm0.21$ & \citet{Ginski_2021} & 088.C-0924(A) \\
 &  &  &  &  & Yes & HWP & 2011-10-31 & NB\_1.64 & 1 & 24 & - & \citet{Ginski_2021} & 088.C-0924(A) \\
 &  &  &  &  & Yes & HWP & 2011-11-01 & Ks & 0.35 & 96 & $1.91\pm0.35$ & \citet{Ginski_2021} & 088.C-0924(A) \\
 &  &  &  &  & Yes & HWP & 2011-11-01 & IB\_2.18 & 0.4 & 60 & - & \citet{Ginski_2021} & 088.C-0924(A) \\ \hline
J05461313-0006048 & V1647 Ori & Orion Var. & - &  & No & WG & 2005-08-05 & H & 1.5 & 32 & - & \citet{Fedele_2007} & 075.C-0489(B) \\
 &  &  &  &  & No & WG & 2005-08-08 & H & 1 & 38 & - & \citet{Fedele_2007} & 075.C-0489(B) \\
 &  &  &  &  & - & WG & 2006-02-26 & Ks & 1.5 & 7 & - & \citet{Fedele_2007} & 075.C-0489(B) \\
 &  &  &  &  & No & WG & 2006-02-26 & Ks & 15 & 28 & - & \citet{Fedele_2007} & 075.C-0489(B) \\
 &  &  &  &  & No & WG & 2006-02-28 & Ks & 25 & 38 & - & \citet{Fedele_2007} & 075.C-0489(B) \\
 &  &  &  &  & - & WG & 2006-03-13 & Ks & 10 & 1 & - & \citet{Fedele_2007} & 075.C-0489(B) \\
 &  &  &  &  & No & WG & 2006-03-13 & Ks & 50 & 47 & - & \citet{Fedele_2007} & 075.C-0489(B) \\ \hline
- & Mon R2 IRS 3 & YSO & - &  & - (Yes) & WG & 2006-12-21 & Ks & 3 & 8 & - & - & 078.C-0554(A) \\ \hline
J06390995+0844097 & R Mon & Herbig Ae/Be & B8$^{(3)}$ &  & Yes & WG & 2006-10-31 & Ks & 0.5 & 64 & $5.11\pm2.07$ & \citet{Murakawa_2008} & 078.C-0554(A) \\
 &  &  &  &  & Yes & WG & 2006-12-21 & Ks & 0.3454 & 64 & $18.90\pm2.13$ & \citet{Murakawa_2008} & 078.C-0554(A) \\ \hline
J07034316-1133062 & Z CMa & Herbig Ae/Be & B5+F5$^{(4)}$ & $5.0^{(5)}$ & Yes & HWP & 2015-01-17 & H & 0.3 & 32 & $<$\,$5.37$ & \citet{Canovas_2015a} & 094.C-0416(A) \\
 &  &  &  &  & Yes & HWP & 2015-01-17 & H & 1 & 8 & - & \citet{Canovas_2015a} & 094.C-0416(A) \\
 &  &  &  &  & Yes & HWP & 2015-01-18 & H & 0.15 & 224 & $<$\,$4.81$ & \citet{Canovas_2015a} & 094.C-0416(A) \\
 &  &  &  &  & Yes & HWP & 2015-01-18 & H & 0.5 & 112 & $<$\,$8.39$ & \citet{Canovas_2015a} & 094.C-0416(A) \\
 &  &  &  &  & Yes & HWP & 2015-01-18 & Ks & 0.15 & 120 & $<$\,$7.21$ & \citet{Canovas_2015a} & 094.C-0416(A) \\
 &  &  &  &  & Yes & HWP & 2015-01-18 & Ks & 0.5 & 96 & $<$\,$9.05$ & \citet{Canovas_2015a} & 094.C-0416(A) \\ \hline
J07192826-4435114 & NX Pup & Herbig Ae/Be & A1$^{(6)}$ &  & No & WG & 2004-12-01 & Ks & 1.5 & 35 & - & - & 074.C-0327(A) \\
 &  &  &  &  & No & WG & 2004-12-01 & H & 2 & 33 & - & - & 074.C-0327(A) \\
 &  &  &  &  & No & WG & 2005-02-07 & Ks & 0.6 & 28 & - & - & 074.C-0327(A) \\
 &  &  &  &  & No & WG & 2005-02-07 & H & 0.7 & 24 & - & - & 074.C-0327(A) \\ \hline
J07503560-3306238 & V646 Pup & Orion Var. & G0-2$^{(7)}$ &  & No & PA & 2008-04-10 & H & 0.35 & 144 & - & - & 381.C-0241(A) \\
 &  &  &  &  & No & PA & 2008-04-10 & H & 2 & 72 & - & - & 381.C-0241(A) \\
 &  &  &  &  & No & PA & 2008-04-10 & H & 5 & 72 & - & - & 381.C-0241(A) \\ \hline
J10590699-7701404 & CR Cha & Orion Var. & K4$^{(8)}$ & $1.9\pm0.2^{(9)}$ & Yes & HWP & 2006-04-09 & Ks & 0.5 & 48 & $<$\,$1.59$ & - & 077.C-0106(A) \\
 &  &  &  &  & Yes & HWP & 2006-04-09 & Ks & 1.2 & 40 & $<$\,$1.37$ & - & 077.C-0106(A) \\
 &  &  &  &  & Yes & HWP & 2006-04-09 & H & 0.35 & 48 & $<$\,$1.80$ & - & 077.C-0106(A) \\
 &  &  &  &  & Yes & HWP & 2006-04-09 & H & 1 & 48 & $<$\,$1.52$ & - & 077.C-0106(A) \\ \hline
J11015191-3442170 & TW Hya & T Tau & M0.5$^{(1)}$ & $0.87^{(10)}$ & No & PA & 2003-02-22 & Ks & 0.5 & 8 & - & \citet{Apai_2004} & 70.C-0482(A) \\
 &  &  &  &  & Yes & PA & 2003-02-22 & Ks & 5 & 8 & $<$\,$17.75$ & \citet{Apai_2004} & 70.C-0482(A) \\
 &  &  &  &  & No & PA & 2003-02-23 & H & 0.5 & 8 & - & \citet{Apai_2004} & 70.C-0482(A) \\
 &  &  &  &  & Yes & PA & 2003-02-23 & H & 5 & 8 & $<$\,$24.68$ & \citet{Apai_2004} & 70.C-0482(A) \\ \hline
J11072074-7738073 & DI Cha & Orion Var. & G2$^{(8)}$ &  & No & HWP & 2006-04-08 & H & 0.5 & 48 & - & - & 077.C-0106(A) \\
 &  &  &  &  & No & HWP & 2006-04-08 & Ks & 0.5 & 32 & - & - & 077.C-0106(A) \\
 &  &  &  &  & No & HWP & 2006-04-08 & NB\_1.64 & 5 & 32 & - & - & 077.C-0106(A) \\
 &  &  &  &  & No & HWP & 2006-04-08 & IB\_2.18 & 1.3 & 36 & - & - & 077.C-0106(A) \\ \hline
J11080329-7739174 & HD 97048 & Herbig Ae/Be & A0$^{(11)}$ & $2.5\pm0.2^{(12)}$ & Yes & HWP & 2006-04-07 & Ks & 0.35 & 32 & $<$\,$6.69$ & \citet{Quanz_2012} & 077.C-0106(A) \\
 &  &  &  &  & Yes & HWP & 2006-04-07 & H & 0.35 & 48 & $0.76\pm0.93$ & \citet{Quanz_2012} & 077.C-0106(A) \\
 &  &  &  &  & Yes & HWP & 2006-04-07 & NB\_1.64 & 3 & 64 & - & \citet{Quanz_2012} & 077.C-0106(A) \\ \hline
J11102788-3731520 & TWA 3A & T Tau & M4.1$^{(1)}$ &  & No & PA & 2003-02-24 & Ks & 0.5 & 8 & - & - & 70.C-0482(A) \\ \hline
J11220530-2446393 & HD 98800 & T Tau & K6.0$^{(1)}$ &  & No & PA & 2003-02-23 & H & 0.5 & 8 & - & - & 70.C-0482(A) \\
 &  &  &  &  & No & PA & 2003-02-23 & H & 1 & 8 & - & - & 70.C-0482(A) \\
 &  &  &  &  & No & PA & 2003-02-23 & Ks & 0.5 & 8 & - & - & 70.C-0482(A) \\
 &  &  &  &  & No & PA & 2003-02-23 & Ks & 1 & 8 & - & - & 70.C-0482(A) \\ \hline
J11315526-3436272 & TWA 5A & T Tau & M2.7$^{(1)}$ &  & No & PA & 2003-02-24 & Ks & 0.5 & 16 & - & - & 70.C-0482(A) \\
 &  &  &  &  & No & PA & 2003-02-24 & H & 0.5 & 16 & - & - & 70.C-0482(A) \\
 &  &  &  &  & No & PA & 2003-02-24 & Ks & 5 & 16 & - & - & 70.C-0482(A) \\
 &  &  &  &  & No & PA & 2003-02-24 & H & 5 & 12 & - & - & 70.C-0482(A) \\ \hline
J11332542-7011412 & HD 100546 & Herbig Ae/Be & A0$^{(13)}$ & $1.9\pm0.1^{(14)}$ & Yes & PA & 2004-06-14 & H & 0.4 & 217 & $3.41\pm0.84$ & - & 073.C-0178(A) \\
 &  &  &  &  & Yes & HWP & 2006-04-06 & Ks & 0.3454 & 52 & $5.24\pm1.44$ & \citet{Quanz_2011} & 077.C-0106(A) \\
 &  &  &  &  & Yes & HWP & 2006-04-06 & H & 0.3454 & 60 & $3.90\pm1.53$ & \citet{Quanz_2011} & 077.C-0106(A) \\
 &  &  &  &  & Yes & HWP & 2006-04-06 & IB\_2.18 & 0.6 & 36 & - & \citet{Quanz_2011} & 077.C-0106(A) \\
 &  &  &  &  & Yes & HWP & 2006-04-06 & NB\_1.64 & 3 & 20 & - & \citet{Quanz_2011} & 077.C-0106(A) \\
 &  &  &  &  & Yes & HWP & 2013-03-30 & Ks & 0.039 & 12 & $2.70\pm2.29$ & \citet{Avenhaus_2014b} & 090.C-0571(B) \\
 &  &  &  &  & Yes & HWP & 2013-03-30 & Ks & 0.3454 & 120 & $5.80\pm1.75$ & \citet{Avenhaus_2014b} & 090.C-0571(B) \\
 &  &  &  &  & No & HWP & 2013-03-30 & Ks & 1.5 & 6 & - & \citet{Avenhaus_2014b} & 090.C-0571(B) \\
 &  &  &  &  & Yes & HWP & 2013-03-30 & H & 0.039 & 16 & $6.07\pm1.81$ & \citet{Avenhaus_2014b} & 090.C-0571(B) \\
 &  &  &  &  & Yes & HWP & 2013-03-30 & H & 0.3454 & 116 & $5.77\pm1.59$ & \citet{Avenhaus_2014b} & 090.C-0571(B) \\
 &  &  &  &  & No & HWP & 2013-03-30 & H & 1.5 & 6 & - & \citet{Avenhaus_2014b} & 090.C-0571(B) \\
 &  &  &  &  & Yes & HWP & 2013-03-30 & L\_prime & 0.175 & 72 & $0.53\pm2.24$ & \citet{Avenhaus_2014b} & 090.C-0571(B) \\
 &  &  &  &  & No & HWP & 2013-03-30 & L\_prime & 2 & 6 & - & \citet{Avenhaus_2014b} & 090.C-0571(B) \\ \hline
J12360103-3952102 & HR 4796 & High-PM (T Tau) & A0$^{(1)}$ & $1.3^{(15)}$ & No & PA & 2003-03-23 & Ks & 0.5 & 16 & - & - & 70.C-0482(A) \\
 &  &  &  &  & No & PA & 2003-03-23 & Ks & 20 & 16 & - & - & 70.C-0482(A) \\
 &  &  &  &  & No & PA & 2003-03-23 & H & 0.35 & 16 & - & - & 70.C-0482(A) \\
 &  &  &  &  & No & PA & 2003-03-23 & H & 5 & 4 & - & - & 70.C-0482(A) \\
 &  &  &  &  & No & PA & 2003-03-23 & H & 15 & 24 & - & - & 70.C-0482(A) \\
 &  &  &  &  & No & PA & 2004-04-06 & H & 3 & 4 & - & - & 073.C-0538(A) \\
 &  &  &  &  & No & PA & 2004-04-06 & H & 10 & 4 & - & - & 073.C-0538(A) \\
 &  &  &  &  & No & PA & 2004-04-06 & H & 60 & 4 & - & - & 073.C-0538(A) \\
 &  &  &  &  & No & PA & 2004-04-06 & IB\_2.21 & 20 & 4 & - & - & 073.C-0538(A) \\
 &  &  &  &  & No & PA & 2004-04-06 & IB\_2.21 & 50 & 4 & - & - & 073.C-0538(A) \\
 &  &  &  &  & No (Yes) & HWP & 2013-04-16 & Ks & 0.35 & 47 & $307.59\pm22.38$ & \citet{Milli_2015} & 091.C-0234(A) \\
 &  &  &  &  & No & HWP & 2013-05-14 & L\_prime & 0.2 & 80 & - & \citet{Milli_2015} & 091.C-0234(A) \\
 &  &  &  &  & No (Yes) & HWP & 2013-05-15 & Ks & 0.5 & 64 & $377.46\pm20.89$ & \citet{Milli_2015} & 091.C-0234(A) \\ \hline
J13220753-6938121 & MP Mus & T Tau & K1$^{(8)}$ &  & No & PA & 2004-05-01 & NB\_1.64 & 2 & 18 & - & - & 073.C-0001(A) \\
 &  &  &  &  & Yes & PA & 2004-05-01 & NB\_1.64 & 3 & 12 & - & - & 073.C-0001(A) \\
 &  &  &  &  & Yes & PA & 2004-05-01 & NB\_1.64 & 5 & 22 & - & - & 073.C-0001(A) \\
 &  &  &  &  & Yes & PA & 2004-05-01 & NB\_1.64 & 10 & 20 & - & - & 073.C-0001(A) \\
 &  &  &  &  & Yes & PA & 2004-05-01 & IB\_2.06 & 1 & 32 & - & - & 073.C-0001(A) \\
 &  &  &  &  & Yes & PA & 2004-05-01 & IB\_2.06 & 3 & 32 & - & - & 073.C-0001(A) \\ \hline
J15154844-3709160 & HD 135344B & YSO & F8$^{(16)}$ & $1.7\pm0.2^{(17)}$ & Yes & HWP & 2012-07-24 & Ks & 0.3454 & 72 & $11.69\pm0.69$ & \citet{Garufi_2013} & 089.C-0611(A) \\
 &  &  &  &  & Yes & HWP & 2012-07-24 & H & 0.5 & 72 & $9.77\pm0.68$ & \citet{Garufi_2013} & 089.C-0611(A) \\
 &  &  &  &  & Yes & HWP & 2012-07-24 & NB\_1.64 & 0.5 & 36 & - & \citet{Garufi_2013} & 089.C-0611(A) \\
 &  &  &  &  & Yes & HWP & 2012-07-24 & NB\_2.17 & 0.5 & 36 & - & \citet{Garufi_2013} & 089.C-0611(A) \\ \hline
J15491210-3539051 & GQ Lup & Orion Var. & K5.0$^{(1)}$ &  & No & HWP & 2012-07-20 & H & 0.15 & 22 & - & - & 089.C-0688(A) \\
 &  &  &  &  & No & HWP & 2012-07-21 & Ks & 0.2 & 23 & - & - & 089.C-0688(A) \\ \hline
J15495775-0355162 & HD 141569 & YSO & A2$^{(13)}$ &  & No & HWP & 2012-07-25 & H & 0.5 & 65 & - & \citet{Garufi_2014} & 089.C-0611(A) \\
 &  &  &  &  & No & HWP & 2012-07-25 & H & 3 & 12 & - & \citet{Garufi_2014} & 089.C-0611(A) \\
 &  &  &  &  & No & HWP & 2012-07-25 & NB\_1.64 & 0.7 & 12 & - & \citet{Garufi_2014} & 089.C-0611(A) \\
 &  &  &  &  & No & HWP & 2012-07-25 & NB\_1.64 & 1 & 12 & - & \citet{Garufi_2014} & 089.C-0611(A) \\
 &  &  &  &  & No & HWP & 2012-07-25 & NB\_1.64 & 2 & 12 & - & \citet{Garufi_2014} & 089.C-0611(A) \\ \hline
J15553378-3709411 & MX Lup & T Tau & K6$^{(18)}$ &  & No & PA & 2003-06-08 & H & 8 & 14 & - & - & 71.C-0507(A) \\ \hline
J15560921-3756057 & IM Lup & Orion Var. & K6.0$^{(1)}$ &  & No & PA & 2003-06-08 & H & 8 & 36 & - & - & 71.C-0507(A) \\ \hline
J15564002-2201400 & HD 142666 & T Tau & F0$^{(13)}$ &  & - & HWP & 2015-07-19 & Ks & 0.5 & 7 & - & - & 60.A-9800(J) \\
 &  &  &  &  & No & HWP & 2015-07-23 & Ks & 0.5 & 80 & - & \citet{Garufi_2017} & 095.C-0658(A) \\ \hline
J15564188-4219232 & HD 142527 & Herbig Ae/Be & A2$^{(19)}$ & $2.2\pm0.3^{(20)}$ & Yes & HWP & 2012-07-18 & H & 0.4 & 12 & $14.46\pm0.59$ & \citet{Canovas_2013} & 089.C-0480(A) \\
 &  &  &  &  & Yes & HWP & 2012-07-18 & H & 1 & 12 & $12.81\pm0.18$ & \citet{Canovas_2013} & 089.C-0480(A) \\
 &  &  &  &  & Yes & HWP & 2012-07-18 & H & 5 & 12 & - & \citet{Canovas_2013} & 089.C-0480(A) \\
 &  &  &  &  & Yes & HWP & 2012-07-23 & Ks & 0.3454 & 72 & $17.12\pm0.15$ & \citet{Quanz_2013} & 089.C-0611(A) \\
 &  &  &  &  & Yes & HWP & 2012-07-23 & H & 0.3454 & 72 & $18.48\pm0.16$ & \citet{Quanz_2013} & 089.C-0611(A) \\
 &  &  &  &  & Yes & HWP & 2012-07-23 & NB\_1.64 & 0.3454 & 24 & - & \citet{Quanz_2013} & 089.C-0611(A) \\
 &  &  &  &  & Yes & HWP & 2012-07-23 & NB\_1.64 & 0.5 & 12 & - & \citet{Quanz_2013} & 089.C-0611(A) \\
 &  &  &  &  & Yes & HWP & 2012-07-23 & NB\_2.17 & 0.3454 & 41 & - & \citet{Quanz_2013} & 089.C-0611(A) \\
 &  &  &  &  & Yes & HWP & 2012-08-11 & Ks & 0.4 & 12 & $12.53\pm0.60$ & \citet{Canovas_2013} & 089.C-0480(A) \\
 &  &  &  &  & Yes & HWP & 2012-08-24 & Ks & 0.4 & 12 & $9.47\pm0.75$ & \citet{Canovas_2013} & 089.C-0480(A) \\
 &  &  &  &  & Yes & HWP & 2012-08-24 & Ks & 4 & 12 & $8.59\pm0.21$ & \citet{Canovas_2013} & 089.C-0480(A) \\ \hline
J16030548-4018254 & EX Lup & Orion Var. & M0$^{(21)}$ &  & No & PA & 2008-04-10 & H & 0.35 & 144 & - & \citet{Kospal_2011} & 381.C-0241(A) \\
 &  &  &  &  & No & PA & 2008-04-10 & H & 2 & 72 & - & \citet{Kospal_2011} & 381.C-0241(A) \\
 &  &  &  &  & No & PA & 2008-04-10 & NB\_1.64 & 2 & 72 & - & \citet{Kospal_2011} & 381.C-0241(A) \\ \hline
J16065795-2743094 & HD 144432 & Herbig Ae/Be & A9$^{(19)}$ &  & No & HWP & 2015-07-22 & Ks & 0.3447 & 1 & - & \citet{Garufi_2017} & 095.C-0658(A) \\
 &  &  &  &  & No & HWP & 2015-07-22 & Ks & 0.345 & 80 & - & \citet{Garufi_2017} & 095.C-0658(A) \\ \hline
J16071159-3903475 & Sz 91 & T Tau & M2.0$^{(1)}$ & $0.58\pm0.07^{(22)}$ & Yes & HWP & 2017-03-20 & Ks & 30 & 44 & $<$\,$9.60$ & \citet{Mauco_2020} & 098.C-0420(A) \\
 &  &  &  &  & Yes & HWP & 2017-03-20 & H & 15 & 112 & $<$\,$4.90$ & \citet{Mauco_2020} & 098.C-0420(A) \\
 &  &  &  &  & No & HWP & 2017-03-20 & H & 20 & 8 & - & \citet{Mauco_2020} & 098.C-0420(A) \\
 &  &  &  &  & - & HWP & 2017-03-20 & Ks & 10 & 1 & - & \citet{Mauco_2020} & 098.C-0420(A) \\ \hline
J16083427-3906181 & V856 Sco & Herbig Ae/Be & M4.6$^{(1)}$ &  & No & HWP & 2015-07-23 & Ks & 0.3454 & 96 & - & \citet{Garufi_2017} & 095.C-0658(A) \\ \hline
J16215769-2429433 & HD 147283 & YSO? & A1$^{(23)}$ &  & No & HWP & 2009-05-01 & Ks & 0.5 & 12 & - & \citet{Murakawa_2012} & 383.D-0197(A) \\
 &  &  &  &  & No & HWP & 2009-05-01 & H & 0.5 & 12 & - & \citet{Murakawa_2012} & 383.D-0197(A) \\ \hline
J16260302-2423360 & Elia 2-14 & Orion Var. & G1$^{(1)}$ &  & No & HWP & 2005-05-31 & H & 0.5 & 12 & - & - & 075.D-0268(A) \\
 &  &  &  &  & No & HWP & 2005-05-31 & Ks & 1 & 12 & - & - & 075.D-0268(A) \\
 &  &  &  &  & No & HWP & 2005-06-01 & H & 32 & 12 & - & - & 075.D-0268(A) \\
 &  &  &  &  & No & HWP & 2005-06-01 & Ks & 45 & 12 & - & - & 075.D-0268(A) \\ \hline
J16262138-2423040 & Elia 2-21 & YSO & - &  & Yes & PA & 2004-04-02 & Ks & 1.789 & 12 & $84.31\pm15.68$ & - & 073.C-0538(A) \\
 &  &  &  &  & Yes & PA & 2004-04-02 & Ks & 20 & 14 & $154.13\pm10.31$ & - & 073.C-0538(A) \\
 &  &  &  &  & Yes & PA & 2004-04-02 & Ks & 60 & 12 & $133.66\pm5.15$ & - & 073.C-0538(A) \\
 &  &  &  &  & Yes & PA & 2004-04-02 & Ks & 120 & 8 & $210.53\pm6.78$ & - & 073.C-0538(A) \\
 &  &  &  &  & - & PA & 2004-04-02 & IB\_2.21 & 200 & 4 & - & - & 073.C-0538(A) \\
 &  &  &  &  & No & PA & 2004-04-05 & H & 30 & 12 & - & - & 073.C-0538(A) \\
 &  &  &  &  & Yes & PA & 2004-04-05 & H & 180 & 12 & $183.17\pm7.09$ & - & 073.C-0538(A) \\
 &  &  &  &  & - & PA & 2004-04-05 & Ks & 10 & 1 & - & - & 073.C-0538(A) \\
 &  &  &  &  & - & PA & 2004-04-05 & Ks & 150 & 4 & - & - & 073.C-0538(A) \\
 &  &  &  &  & Yes & PA & 2004-04-05 & L\_prime & 0.175 & 38 & $91.35\pm11.46$ & - & 073.C-0538(A) \\ \hline
J16262803-2526477 & ROXs 12 & YSO & M0.0$^{(24)}$ &  & - & HWP & 2016-06-13 & Ks & 6 & 1 & - & - & 097.C-0644(B) \\
 &  &  &  &  & No & HWP & 2016-06-13 & Ks & 15 & 98 & - & - & 097.C-0644(B) \\
 &  &  &  &  & No & HWP & 2018-03-09 & Ks & 2 & 8 & - & - & 0100.C-0492(C) \\
 &  &  &  &  & No & HWP & 2018-03-09 & Ks & 55 & 36 & - & - & 0100.C-0492(C) \\ \hline
J16263416-2423282 & Elia 2-25 & T Tau & B3$^{(25)}$ &  & Yes & PA & 2003-06-18 & Ks & 0.109 & 10 & $12.64\pm1.25$ & - & 60.A-9026(A) \\
 &  &  &  &  & - & PA & 2003-06-18 & Ks & 0.345 & 5 & - & - & 60.A-9026(A) \\
 &  &  &  &  & Yes & PA & 2003-06-18 & Ks & 1 & 6 & $7.91\pm5.47$ & - & 60.A-9026(A) \\
 &  &  &  &  & Yes & PA & 2003-06-18 & Ks & 15 & 6 & $9.13\pm2.50$ & - & 60.A-9026(A) \\
 &  &  &  &  & No & PA & 2003-06-18 & H & 1 & 6 & - & - & 60.A-9026(A) \\
 &  &  &  &  & No & HWP & 2018-05-31\textdagger & H & 0.3447 & 10 & - & \citet{Millar_Blanchaer_2020} & 0101.C-0561(B) \\
 &  &  &  &  & No & HWP & 2018-06-12\textdagger & H & 0.35 & 10 & - & \citet{Millar_Blanchaer_2020} & 0101.C-0561(B) \\
 &  &  &  &  & No & HWP & 2019-04-29\textdagger & H & 1 & 9 & - & - & 0103.C-0728(A) \\
 &  &  &  &  & No & HWP & 2019-04-30\textdagger & H & 2 & 16 & - & - & 0103.C-0728(A) \\
 &  &  &  &  & No & HWP & 2019-05-01\textdagger & Ks & 1.5 & 16 & - & - & 0103.C-0728(A) \\
 &  &  &  &  & No & HWP & 2019-05-03\textdagger & H & 0.5 & 16 & - & - & 0103.C-0728(A) \\ \hline
J16270677-2438149 & WL 17 & YSO & - &  & No & PA & 2004-04-04 & Ks & 12 & 12 & - & - & 073.C-0538(A) \\
 &  &  &  &  & No & PA & 2004-04-04 & Ks & 60 & 12 & - & - & 073.C-0538(A) \\
 &  &  &  &  & No & PA & 2004-04-04 & Ks & 120 & 12 & - & - & 073.C-0538(A) \\
 &  &  &  &  & No & PA & 2004-04-04 & IB\_2.21 & 180 & 6 & - & - & 073.C-0538(A) \\ \hline
J16270943-2437187 & Elia 2-29 & YSO & - &  & - & PA & 2004-04-01 & H & 5 & 1 & - & \citet{Huelamo_2007} & 073.C-0538(A) \\
 &  &  &  &  & Yes & PA & 2004-04-01 & H & 20 & 14 & $43.34\pm11.58$ & \citet{Huelamo_2007} & 073.C-0538(A) \\
 &  &  &  &  & Yes & PA & 2004-04-01 & H & 60 & 13 & $95.23\pm6.08$ & \citet{Huelamo_2007} & 073.C-0538(A) \\
 &  &  &  &  & Yes & PA & 2004-04-01 & H & 120 & 13 & $81.80\pm14.60$ & \citet{Huelamo_2007} & 073.C-0538(A) \\
 &  &  &  &  & Yes & PA & 2004-04-01 & Ks & 1.789 & 22 & $48.43\pm4.40$ & \citet{Huelamo_2007} & 073.C-0538(A) \\
 &  &  &  &  & Yes & PA & 2004-04-01 & Ks & 10 & 20 & $61.09\pm3.74$ & \citet{Huelamo_2007} & 073.C-0538(A) \\
 &  &  &  &  & Yes & PA & 2004-04-01 & Ks & 20 & 14 & $62.70\pm1.39$ & \citet{Huelamo_2007} & 073.C-0538(A) \\
 &  &  &  &  & Yes & PA & 2004-04-01 & IB\_2.21 & 5 & 12 & - & \citet{Huelamo_2007} & 073.C-0538(A) \\
 &  &  &  &  & - & PA & 2004-04-01 & IB\_2.21 & 6 & 1 & - & \citet{Huelamo_2007} & 073.C-0538(A) \\
 &  &  &  &  & Yes & PA & 2004-04-02 & Ks & 1.789 & 13 & $70.90\pm4.90$ & \citet{Huelamo_2007} & 073.C-0538(A) \\
 &  &  &  &  & Yes & PA & 2004-04-02 & Ks & 60 & 14 & $88.68\pm1.80$ & \citet{Huelamo_2007} & 073.C-0538(A) \\
 &  &  &  &  & No & PA & 2004-04-03 & L\_prime & 0.18 & 15 & - & \citet{Huelamo_2007} & 073.C-0538(A) \\
 &  &  &  &  & No & PA & 2004-04-06 & NB\_1.64 & 120 & 4 & - & \citet{Huelamo_2007} & 073.C-0538(A) \\
 &  &  &  &  & No & PA & 2004-04-06 & NB\_1.64 & 240 & 4 & - & \citet{Huelamo_2007} & 073.C-0538(A) \\
 &  &  &  &  & Yes & PA & 2004-04-06 & IB\_2.21 & 2 & 8 & - & \citet{Huelamo_2007} & 073.C-0538(A) \\
 &  &  &  &  & Yes & PA & 2004-04-06 & IB\_2.21 & 10 & 4 & - & \citet{Huelamo_2007} & 073.C-0538(A) \\
 &  &  &  &  & Yes & PA & 2004-04-06 & IB\_2.21 & 60 & 4 & - & \citet{Huelamo_2007} & 073.C-0538(A) \\
 &  &  &  &  & Yes & PA & 2004-04-06 & NB\_2.12 & 3 & 4 & - & \citet{Huelamo_2007} & 073.C-0538(A) \\
 &  &  &  &  & Yes & PA & 2004-04-06 & NB\_2.12 & 6 & 4 & - & \citet{Huelamo_2007} & 073.C-0538(A) \\
 &  &  &  &  & Yes & PA & 2004-04-06 & NB\_2.12 & 20 & 4 & - & \citet{Huelamo_2007} & 073.C-0538(A) \\
 &  &  &  &  & Yes & PA & 2004-04-06 & NB\_2.12 & 100 & 4 & - & \citet{Huelamo_2007} & 073.C-0538(A) \\
 &  &  &  &  & Yes & PA & 2004-04-06 & NB\_3.74 & 1 & 8 & - & \citet{Huelamo_2007} & 073.C-0538(A) \\ \hline
J16271569-2438434 & WL 20 & YSO & - &  & No & PA & 2004-04-03 & Ks & 12 & 8 & - & - & 073.C-0538(A) \\
 &  &  &  &  & No & PA & 2004-04-03 & Ks & 25 & 6 & - & - & 073.C-0538(A) \\
 &  &  &  &  & - & PA & 2004-04-03 & Ks & 60 & 3 & - & - & 073.C-0538(A) \\
 &  &  &  &  & No & PA & 2004-04-03 & Ks & 120 & 7 & - & - & 073.C-0538(A) \\
 &  &  &  &  & No & PA & 2004-04-03 & Ks & 240 & 4 & - & - & 073.C-0538(A) \\ \hline
J16271951-2441403 & EM* SR 12 & Orion Var. & M0$^{(26)}$ &  & No & HWP & 2018-03-11 & Ks & 4 & 8 & - & - & 0100.C-0492(D) \\
 &  &  &  &  & No & HWP & 2018-03-11 & Ks & 55 & 24 & - & - & 0100.C-0492(D) \\ \hline
J16272461-2441034 & CRBR 2422.8-3423 & YSO & - &  & - & PA & 2004-04-05 & Ks & 60 & 1 & - & - & 073.C-0538(A) \\
 &  &  &  &  & - & PA & 2004-04-05 & Ks & 150 & 2 & - & - & 073.C-0538(A) \\
 &  &  &  &  & - & PA & 2004-04-05 & Ks & 180 & 3 & - & - & 073.C-0538(A) \\ \hline
J16272693-2440508 & YLW 15 & YSO & K5$^{(27)}$ &  & No & PA & 2004-04-04 & Ks & 12 & 14 & - & - & 073.C-0538(A) \\
 &  &  &  &  & No & PA & 2004-04-04 & Ks & 200 & 12 & - & - & 073.C-0538(A) \\
 &  &  &  &  & No & PA & 2004-04-04 & IB\_2.21 & 100 & 9 & - & - & 073.C-0538(A) \\ \hline
J16272802-2439335 & YLW 16A & YSO & - &  & Yes & PA & 2004-04-03 & Ks & 60 & 13 & $401.91\pm8.65$ &  & 073.C-0538(A) \\
 &  &  &  &  & Yes & PA & 2004-04-03 & Ks & 200 & 12 & $383.24\pm9.41$ &  & 073.C-0538(A) \\
 &  &  &  &  & - & PA & 2004-04-03 & Ks & 400 & 1 & - &  & 073.C-0538(A) \\ \hline
J16275209-2440503 & ROXs 31 & T Tau & K7.5$^{(25)}$ &  & No & HWP & 2018-03-10 & Ks & 2 & 8 & - & - & 0100.C-0492(D) \\ \hline
J16311431-2434150 & ROXs 42A & T Tau & F/G$^{(28)}$ &  & No & HWP & 2018-05-26\textdagger & Ks & 55 & 28 & - & - & 0100.C-0492(F) \\ \hline
J16311501-2432436 & ROXs 42B & T Tau & M0$^{(28)}$ &  & No & HWP & 2018-03-23 & Ks & 40 & 24 & - & - & 0100.C-0492(E) \\
 &  &  &  &  & - & HWP & 2018-03-23 & Ks & 50 & 2 & - & - & 0100.C-0492(E) \\
 &  &  &  &  & - & HWP & 2018-03-23 & Ks & 60 & 2 & - & - & 0100.C-0492(E) \\ \hline
J16311574-2434022 & ROXs 42C & Orion Var. & K6$^{(28)}$ &  & No & HWP & 2018-03-23 & Ks & 0.3447 & 8 & - & - & 0100.C-0492(E) \\ \hline
J16323219-4455306 & V346 Nor & Orion Var. & - &  & No & PA & 2008-04-10 & H & 1 & 72 & - & \citet{Kospal_2017} & 381.C-0241(A) \\
 &  &  &  &  & No & PA & 2008-04-10 & H & 2 & 72 & - & \citet{Kospal_2017} & 381.C-0241(A) \\
 &  &  &  &  & No & PA & 2008-04-10 & H & 10 & 72 & - & \citet{Kospal_2017} & 381.C-0241(A) \\
 &  &  &  &  & No & PA & 2008-04-10 & H & 20 & 72 & - & \citet{Kospal_2017} & 381.C-0241(A) \\ \hline
J16401792-2353452 & HD 150193 & Herbig Ae/Be & B9.5$^{(29)}$ &  & No & HWP & 2007-06-04 & H & 0.35 & 8 & - & - & 079.C-0189(A) \\
 &  &  &  &  & No & HWP & 2012-07-24 & H & 0.3454 & 96 & - & \citet{Garufi_2014} & 089.C-0611(A) \\
 &  &  &  &  & No & HWP & 2012-07-24 & NB\_1.64 & 0.3454 & 36 & - & \citet{Garufi_2014} & 089.C-0611(A) \\
 &  &  &  &  & No & HWP & 2012-07-25 & Ks & 0.5 & 49 & - & \citet{Garufi_2014} & 089.C-0611(A) \\ \hline
J16544485-3653185 & AK Sco & Herbig Ae/Be & F5$^{(19)}$ & $1.35\pm0.07^{(30)}$ & Yes & HWP & 2015-07-22 & Ks & 2 & 96 & $<$\,$3.41$ & \citet{Garufi_2017} & 095.C-0658(A) \\ \hline
J17310584-3508292 & HD 319896 & Herbig Ae/Be? & B4$^{(31)}$ &  & No & HWP & 2005-06-01 & H & 15 & 12 & - & - & 075.D-0268(A) \\
 &  &  &  &  & No & HWP & 2005-06-01 & Ks & 15 & 12 & - & - & 075.D-0268(A) \\
 &  &  &  &  & No & HWP & 2005-06-02 & H & 10 & 12 & - & - & 075.D-0268(A) \\ \hline
J17562128-2157218 & HD 163296 & Herbig Ae/Be & A1$^{(3)}$ & $2.23\pm0.22^{(32)}$ & Yes & HWP & 2012-07-23 & H & 0.3454 & 72 & $0.89\pm0.26$ & \citet{Garufi_2014} & 089.C-0611(A) \\
 &  &  &  &  & Yes & HWP & 2012-07-23 & Ks & 0.3454 & 36 & $1.25\pm0.08$ & \citet{Garufi_2014} & 089.C-0611(A) \\
 &  &  &  &  & No & HWP & 2012-07-23 & NB\_1.64 & 0.3454 & 36 & - & \citet{Garufi_2014} & 089.C-0611(A) \\
 &  &  &  &  & No & HWP & 2012-07-23 & NB\_2.17 & 0.3454 & 12 & - & \citet{Garufi_2014} & 089.C-0611(A) \\
 &  &  &  &  & No & HWP & 2015-07-22 & Ks & 0.3454 & 80 & - & \citet{Garufi_2017} & 095.C-0658(A) \\ \hline
J18143956-1752023 & W 33a & YSO & - &  & No & HWP & 2010-09-28 & H & 120 & 16 & - & - & 385.C-0301(A) \\ \hline
J18242978-2946492 & HD 169142 & Herbig Ae/Be & F1$^{(13)}$ & $1.79^{(33)}$ & Yes & HWP & 2007-06-04 & Ks & 0.35 & 36 & $<$\,$6.96$ & - & 079.C-0189(A) \\
 &  &  &  &  & Yes & HWP & 2007-06-04 & Ks & 5 & 78 & $<$\,$8.66$ & - & 079.C-0189(A) \\
 &  &  &  &  & Yes & HWP & 2007-06-04 & Ks & 10 & 24 & $<$\,$9.07$ & - & 079.C-0189(A) \\
 &  &  &  &  & Yes & HWP & 2007-06-04 & H & 0.4 & 20 & $3.07\pm1.09$ & - & 079.C-0189(A) \\
 &  &  &  &  & Yes & HWP & 2007-06-04 & H & 4 & 36 & $3.73\pm0.36$ & - & 079.C-0189(A) \\
 &  &  &  &  & Yes & HWP & 2007-06-04 & H & 10 & 44 & - & - & 079.C-0189(A) \\
 &  &  &  &  & Yes & HWP & 2012-05-04 & H & 0.4 & 12 & $2.62\pm2.96$ & - & 089.C-0480(A) \\
 &  &  &  &  & Yes & HWP & 2012-05-20 & Ks & 20 & 7 & $<$\,$19.37$ & - & 089.C-0480(A) \\
 &  &  &  &  & Yes & HWP & 2012-07-25 & H & 1 & 48 & $2.64\pm0.65$ & \citet{Quanz_2013} & 089.C-0611(A) \\
 &  &  &  &  & Yes & HWP & 2012-07-25 & NB\_1.64 & 1 & 24 & - & \citet{Quanz_2013} & 089.C-0611(A) \\
 &  &  &  &  & Yes & HWP & 2012-08-11 & Ks & 0.4 & 12 & $<$\,$14.91$ & - & 089.C-0480(A) \\
 &  &  &  &  & Yes & HWP & 2012-08-24 & Ks & 0.4 & 12 & $4.43\pm3.53$ & - & 089.C-0480(A) \\
 &  &  &  &  & Yes & HWP & 2012-08-24 & Ks & 20 & 12 & - & - & 089.C-0480(A) \\
 &  &  &  &  & Yes & HWP & 2012-08-25 & H & 10 & 12 & $3.24\pm0.18$ & - & 089.C-0480(A) \\ \hline
J19005804-3645048 & - & YSO & M0.75$^{(34)}$ &  & No & HWP & 2019-06-07\textdagger & H & 3 & 56 & - & \citet{Christiaens_2021} & 0103.C-0865(A) \\
 &  &  &  &  & - & HWP & 2019-06-07\textdagger & H & 0.8 & 4 & - & \citet{Christiaens_2021} & 0103.C-0865(A) \\ \hline
J19015367-3657081 & R CrA & Herbig Ae/Be & B5$^{(35)}$ & $3.02\pm0.43^{(36)}$ & Yes & HWP & 2012-07-18 & H & 0.5 & 12 & $<$\,$35.76$ & - & 089.C-0480(A) \\ \hline
J19290085+0938429 & Parsamian 21 & Orion Var. & F5$^{(37)}$ &  & Yes & PA & 2004-06-17 & H & 10 & 72 & $399.65\pm2.43$ & \citet{Kospal_2008} & 073.C-0721(A) \\
 &  &  &  &  & Yes & PA & 2004-06-17 & H & 80 & 24 & $<$\,$366.48$ & \citet{Kospal_2008} & 073.C-0721(A)

\end{longtable}
}
}
{\raggedright\footnotesize
    \textbf{Notes. }
    (a) The abbreviations of the SIMBAD object types are: Orion Var. for Orion variable stars; Herbig Ae/Be for Herbig Ae stars; T Tau for T Tauri stars; Herbig Ae/Be for Herbig Be stars; High-PM for high-proper motion stars; and YSO for young stellar objects. Abbreviations followed by a question mark are candidate object types and those listed in parentheses show previous identifications. \\
    (b) Datasets where the cross-correlation of Sect. \ref{sect:cross_corr} could not be applied, due to incomplete coverage of both $Q$ and $U$, present a hyphen (-) in the `Detection' column. Instances where the cross-correlation analysis resulted in a non-detection despite clear signs of polarised light from a visual inspection are appended with `(Yes)'. \\
    (c) Datasets indicated with \textdagger\ were observed after April 11, 2018, when the HWP rotation mechanism failed \citep{Millar_Blanchaer_2020}. After its repair, the motor encoder position no longer corresponded to the same polarisation angle. PIPPIN is currently not equipped to correct for this systematic offset in the observed polarisation angle, and results from these datasets should therefore not be trusted.
    
    \textbf{References. }
    (1)~\citet{Herczeg_2014};
    (2)~\citet{Ginski_2021};
    (3)~\citet{Mora_2001};
    (4)~\citet{Covino_1984};
    (5)~\citet{Millan_2002};
    (6)~\citet{Skiff_2014};
    (7)~\citet{Reipurth_2002};
    (8)~\citet{Torres_2006};
    (9)~\citet{Hussain_2009};
    (10)~\citet{van_Boekel_2017};
    (11)~\citet{Irvine_1977};
    (12)~\citet{van_den_Ancker_1998};
    (13)~\citet{Gray_2017};
    (14)~\citet{Fairlamb_2015};
    (15)~\citet{Olofsson_2019};
    (16)~\citet{Coulson_1995};
    (17)~\citet{Mueller_2011};
    (18)~\citet{Krautter_1997};
    (19)~\citet{Houk_1982};
    (20)~\citet{Verhoeff_2011};
    (21)~\citet{Alcala_2017};
    (22)~\citet{Mauco_2020};
    (23)~\citet{Houk_1988};
    (24)~\citet{Rizzuto_2015};
    (25)~\citet{Wilking_2005};
    (26)~\citet{Pecaut_2016};
    (27)~\citet{Greene_2002};
    (28)~\citet{Bouvier_1992};
    (29)~\citet{Levenhagen_2006};
    (30)~\citet{Alencar_2003};
    (31)~\citet{Vieira_2003};
    (32)~\citet{Alecian_2013};
    (33)~\citet{Blondel_2006};
    (34)~\citet{Romero_2012};
    (35)~\citet{Gray_2006};
    (36)~\citet{Sissa_2019};
    (37)~\citet{Staude_1992} \\
}
\end{landscape}
\clearpage
    
\section{PIPPIN configuration keywords} \label{app:keywords}
\begin{table*}[h!]
\centering
\caption{Keywords and values recognised by PIPPIN in the configuration file. \label{tab:PIPPIN_keywords}}

{\setlength{\tabcolsep}{3.5pt}
{\footnotesize
\begin{tabular}{lll}
\hline\hline
\textbf{PIPPIN configuration keywords} & \textbf{Recognised values} & \textbf{Description} \\
\hline\hline
& & \\
\textbf{Pre-processing options} & & \\
\hline
\texttt{run\_pre\_processing} & \texttt{bool} & Set to \texttt{False} to only run PDI functions (\texttt{True}) \\
\texttt{remove\_data\_products} & \texttt{bool} & Remove reduced and sky-subtraction images (\texttt{True}) \\
\texttt{split\_observing\_blocks} & \texttt{bool} & Classification by observing ID (\texttt{True}) \\
\texttt{y\_pixel\_range} & \texttt{[int,int]} & Image cropping for more efficient reduction (\texttt{[0,1024]}) \\
& & \\
\textbf{Sky-subtraction} & & \\
\hline
\texttt{sky\_subtraction\_method} & [\texttt{dithering-offset}, \texttt{box-median}] & Sky-subtraction method (\texttt{dithering-offset}) \\
\texttt{sky\_subtraction\_min\_offset} & \texttt{int} & Minimum pixel offset between dithering positions or \\
 & & box-median regions (\texttt{100}) \\
\texttt{remove\_horizontal\_stripes} & \texttt{bool} & Remove read-out pattern with more aggressive gradient \\
 & & fitting (\texttt{False}) \\
& & \\
\textbf{Centering} & & \\
\hline
\texttt{centering\_method} & [\texttt{single-Moffat}, \texttt{double-Moffat}, \texttt{maximum}] & Beam-fitting method (\texttt{single-Moffat}) \\
\texttt{tied\_offset} & \texttt{bool} & Constrain the beam separation (\texttt{False}) \\
& & \\
\textbf{PDI options} & & \\
\hline
\texttt{size\_to\_crop} & \texttt{[int,int]} & Height and width of final data products (\texttt{[120,120]}) \\
\texttt{r\_inner\_IPS} & \texttt{[int,...]} & Inner annulus radius for $IP$-subtraction (\texttt{[3,6,9]}) \\
\texttt{r\_outer\_IPS} & \texttt{[int,...]} & Outer annulus radius for $IP$-subtraction (\texttt{[6,9,12]}) \\
\texttt{crosstalk\_correction} & \texttt{bool} & Correct for reduced $U$ efficiency (\texttt{False}) \\
\texttt{minimise\_U\_phi} & \texttt{bool} & Minimise the $U_\phi$ (\texttt{False}) \\
\texttt{r\_crosstalk} & \texttt{[int,int]} & Inner and outer annulus radii to use for crosstalk- \\
 & & correction and $U_\phi$ minimisation (\texttt{[7,17]}) \\
& & \\
\textbf{Object information} & & \\
\hline
\texttt{object\_name} & \texttt{str} & Object identifier in SIMBAD (\textit{derived from } \\
 & & \textit{directory-name}) \\
\texttt{disk\_pos\_angle} & \texttt{float} & Disk position-angle in degrees (\texttt{0.0}) \\
\texttt{disk\_inclination} & \texttt{float} & Disk inclination in degrees (\texttt{0.0}) \\
\hline
\end{tabular}
}
}
\end{table*}
\clearpage
    
\section{Extended data products of Parsamian 21 and Elia 2-21} \label{app:supp_figures}
\begin{figure*}[h]
    \centering
    \includegraphics[width=14cm]{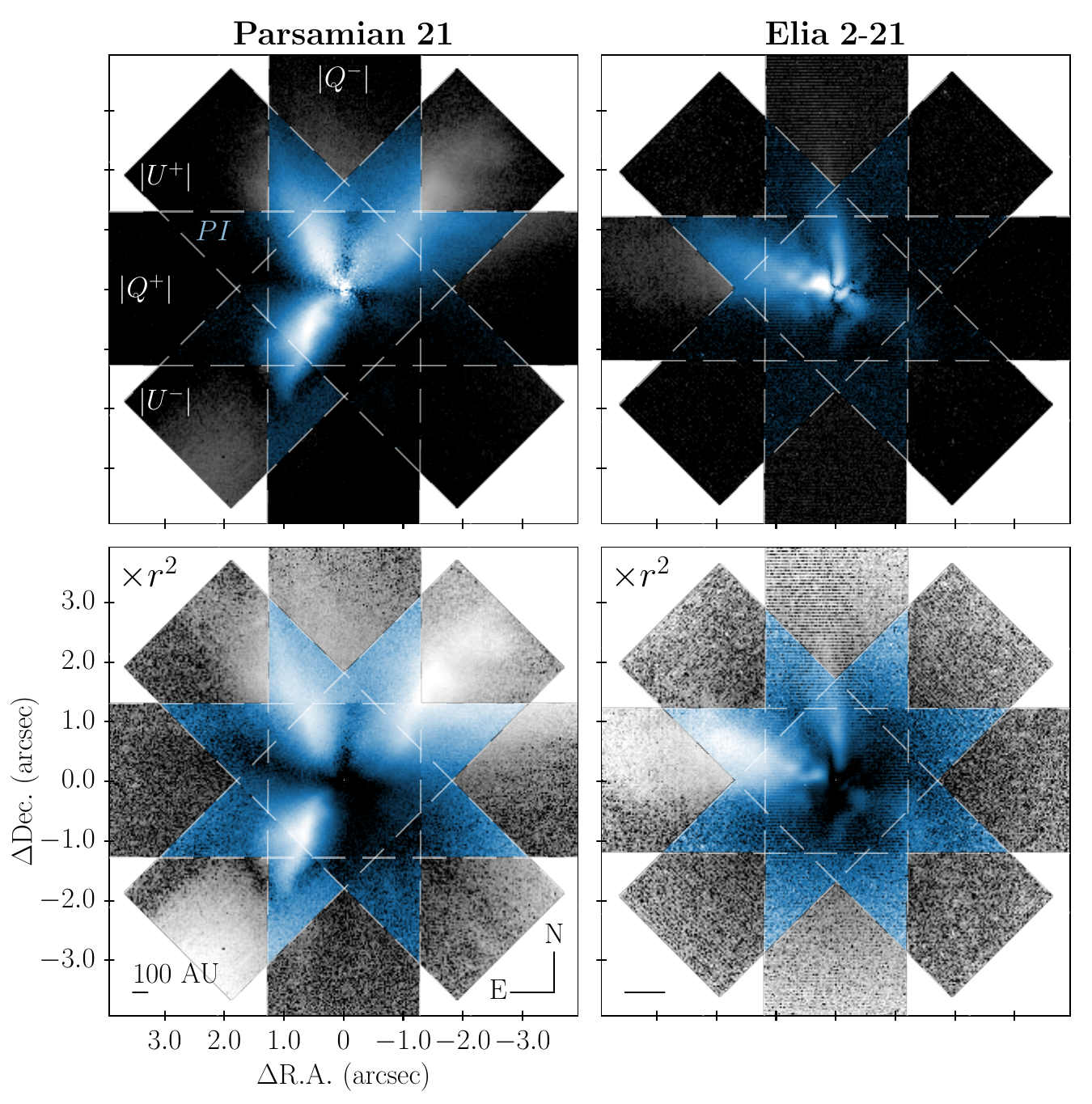}
    \caption{Polarised light for the embedded YSOs Parsamian 21 (\textit{left panels}) and Elia 2-21 (\textit{right panels}). The \textit{top panels} show the polarised intensity, $PI,$ with a blue colour map, while the grey colours display the absolute values of the linear Stokes components $|Q^\pm|$ and $|U^\pm|$. In the \textit{bottom panels}, these values are scaled by the squared separation from the centre. The dashed lines delineate the sections of the sky observed by one of the components. These sections overlap in the centre, resulting in an eight-pointed star, where the polarised intensity image can be computed as $Q$ and $U$ are both covered. The south-eastern region of the $U^-$ observation of Parsamian 21 is contaminated with signal from another dithering position, introduced during the sky-subtraction.}
    \label{fig:gallery_embedded_with_QU}
\end{figure*}
        
\end{appendix}

\end{document}